\documentclass{article}

\usepackage{arxiv}

\usepackage[utf8]{inputenc} % allow utf-8 input
\usepackage[T1]{fontenc}    % use 8-bit T1 fonts
\usepackage{hyperref}       % hyperlinks
\usepackage{url}            % simple URL typesetting
\usepackage{booktabs}       % professional-quality tables
\usepackage{amsfonts}       % blackboard math symbols
\usepackage{nicefrac}       % compact symbols for 1/2, etc.
\usepackage{microtype}      % microtypography
\usepackage{lipsum}		% Can be removed after putting your text content
\usepackage{graphicx}
\usepackage{natbib}
\usepackage{doi}

\title{Biometrics in Extended Reality: A  Review}

\author{
 Ayush Agarwal \\
  Department of Electrical Engineering\\
  Indian Institute of Technology Dharwad\\
  Karnataka, India, Pin Code: 580011 \\
  \texttt{201081001@iitdh.ac.in} \\
   \And
 Raghavendra Ramachandra \\
 Department of Information Security and Communication Technology\\
  Norwegian University of Science and Technology (NTNU)\\  
  Gjøvik, Norway \\
  \texttt{raghavendra.ramachandra@ntnu.no} \\  
   \And
Sushma Venkatesh \\
  AiBA AS\\
  Norway\\
  \texttt{sushma@aiba.ai}\\
  \And
S. R. Mahadeva Prasanna \\
Department of Electrical Engineering\\
Indian Institute of Technology Dharwad\\
Karnataka, India, Pin Code: 580011 \\
\texttt{prasanna@iitdh.ac.in} \\
}
\renewcommand{\shorttitle}{Biometrics in Extended Reality: A  Review}

\hypersetup{
pdftitle={A template for the arxiv style},
pdfsubject={q-bio.NC, q-bio.QM},
pdfauthor={David S.~Hippocampus, Elias D.~Striatum},
pdfkeywords={First keyword, Second keyword, More},
}

\begin{document}
\maketitle

\begin{abstract}
In the domain of Extended Reality (XR), particularly Virtual Reality (VR), extensive research has been devoted to harnessing this transformative technology in various real-world applications. However, a critical challenge that must be addressed before unleashing the full potential of XR in practical scenarios is to ensure robust security and safeguard user privacy.  This paper presents a systematic survey of the utility of biometric characteristics applied in  the XR environment. To this end, we present a comprehensive overview of the different types of biometric modalities used for authentication and representation of users in a virtual environment. We discuss different biometric vulnerability gateways in general XR systems for the first time in the literature along with taxonomy. A comprehensive discussion on generating and authenticating biometric-based photorealistic avatars in XR environments is presented with a stringent taxonomy. We also discuss the availability of different datasets that are widely employed in evaluating biometric authentication in XR environments together with performance evaluation metrics. Finally, we discuss the open challenges and potential future work that need to be addressed in the field of biometrics in XR.
\end{abstract}

% keywords can be removed
\keywords{Biometrics \and Authentication \and Extended reality \and Attacks \and Virtual Reality \and Security \and Avatars \and Synthetic AI}

\section{Introduction}

\label{Sec:Intro}
Extended Reality (XR) is an umbrella term that encompasses various immersive technologies, including Virtual Reality (VR), Augmented Reality (AR), and Mixed Reality (MR). XR combines real and virtual environments to create a blended experience that allows users to interact with digital content and the physical world. VR completely immerses users in a simulated environment, AR overlays digital content onto the real world, and MR blends digital and real-world elements to create interactive and dynamic experiences. The market value of XR has increased rapidly in recent years from USD 6.1 billion in 2016 to USD 42.83 billion in 2022. It is expected to grow further, reaching USD 345.09 billion by 2030\footnote{https://www.vantagemarketresearch.com/}.  This growth is driven by the increasing adoption of XR in various sectors as well as advancements in XR technology and hardware.

The evolution of AI has the potential to significantly enhance the usability of extended reality (XR) by making interactions within virtual environments more seamless and accessible. Through AI-powered natural language processing, users can rely on improved voice commands and real-time translations, enabling intuitive communication across global audiences. AI can also personalize XR experiences through adaptive learning systems, tailoring environments, and tutorials to meet individual needs, thereby reducing the learning curve. In addition, AI-driven dynamic content generation fosters more immersive and interactive virtual worlds, creating a richer experience for users. Collaboration within XR can be enhanced as AI facilitates the real-time tracking of gestures and facial expressions, enhancing smoother communication in shared virtual spaces.

\begin{figure*}[]
\centering
\includegraphics[height= 250pt,width= 350pt]{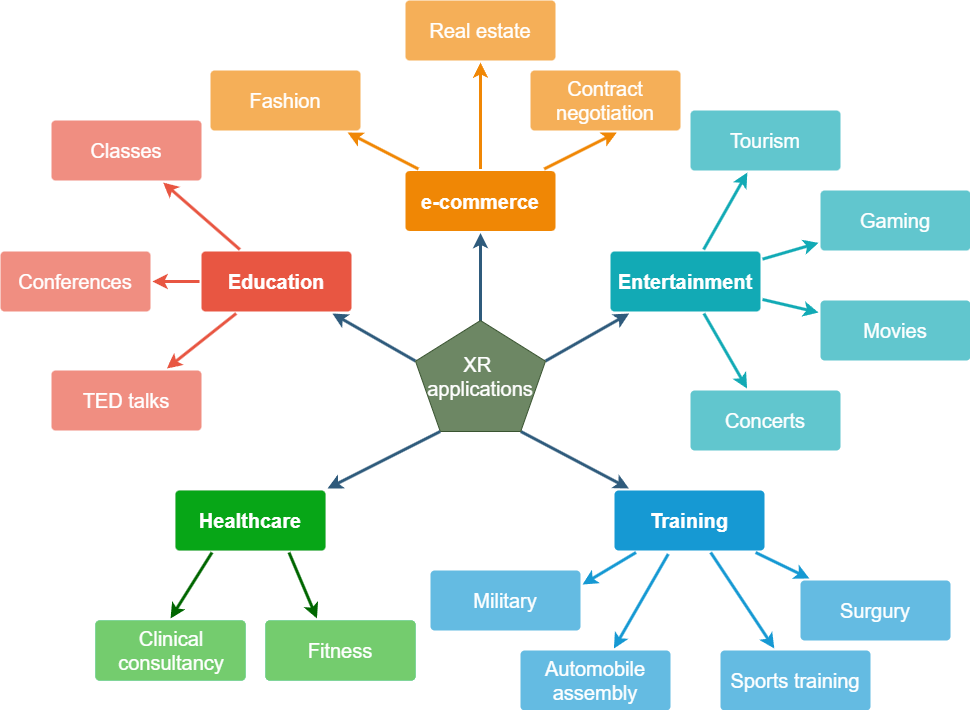}
\caption{Applications of extended reality (XR) systems in different sectors.}
\label{VR_applications}
\end{figure*}

Extended reality (XR)  systems are widely utilized in several critical applications, including healthcare, education, and the military, in addition to societal applications such as healthcare, tourism, entertainment, sports, and e-commerce[~\cite{sharma2017virtual}]. In healthcare, XR is used for medical training, surgery simulation, patient education, and therapy for conditions such as anxiety disorders and chronic pain. In education, XR is used to create immersive learning experiences and simulations that enhance student engagement and understanding of complex concepts. In tourism, XR is used to create virtual tours of famous landmarks and tourist destinations, providing users with a unique and immersive experience. In entertainment, XR is used to create interactive and immersive gaming experiences that push the boundaries of traditional gaming. The military and defense sectors also use XR for training and simulation. XR simulations are used to train soldiers in combat situations, simulate vehicle and aircraft operations, and create virtual battlefields. E-commerce also uses XR to create virtual shopping experiences, allowing customers to browse and purchase products in virtual stores. Sports is another sector where XR is used to enhance training and performance analysis. XR simulations are used to train athletes in various sports, including football, basketball, and golf, to improve their technique and decision-making skills. XR is also used for sports broadcasting, providing viewers with an immersive and interactive experience. The detailed overview of XR systems applications is presented in Figure \ref{VR_applications}.

\begin{figure}[]
\centering
\includegraphics[height= 200pt,width= 300pt]{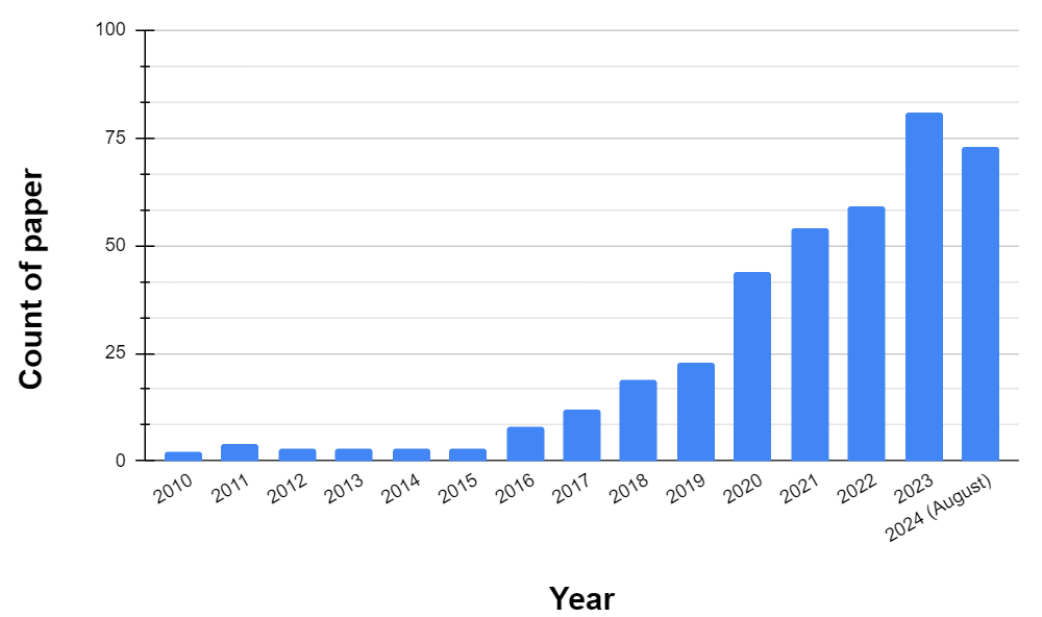}
\caption{The number of year-wise works in the direction of security and privacy in VR applications indicates the increasing interest of researchers.}
\label{fig:VR_countVsYear}
\end{figure}

The use of XR in critical applications has raised concerns regarding the security and trustworthiness of such systems. Evolving technology with XR has resulted in semi-autonomous avatars that have increased naturalness (or realism) in human-to-human interactions. At present, the interaction in XR goes beyond the mere exchange of words, where dynamic support for nonverbal cues and photorealistic avatars is possible. Therefore, a highly personalized representation of user behaviors with photorealistic rendering will increase the trustworthiness of individuals involved in the XR environment.  Furthermore, the development of XR systems in various fields has provided a huge portal for the flow of user data for fraudulent activities by people with malicious intentions. Data piracy and theft in a virtual world can be used to unlock real-world security systems. Therefore, there is a need for a security and privacy system that protects users’ information generated in the virtual world from affecting their information in the real world.

The wide application of XR in critical applications mandates appropriate access control when using XR tools, and also ensures the trustworthiness of the digital person representation in XR space, which features biometric characteristics. All existing XR devices utilize both behavioral and physiological biometric characteristics, including eye tracking (e.g., Vive Pro Eye, Varjo, Neo 2 Eye, Meta), heart rate (e.g., Teslasuit), facial expressions (e.g., HTC Vive), and hand movements (e.g., Manus VR). Popular (e.g., Meta) and upcoming (e.g., Apple) VR headsets in the market use iris recognition to improve their trustworthiness and security of VR headsets. On the other hand, augmenting the capabilities of users to experience seamless and persistent physical-digital experiences in virtual environments based on biometric characteristics is extensively studied. The seamless integration of a user’s personal appearance and behavioral style results in a photorealistic full-body avatar[~\cite{lattas2023fitme}]. Therefore, the created avatar must be protected against identity theft and impersonation attacks.

\begin{table*}
\centering
\caption{Comparison of existing survey papers with this survey paper. The main emphasis of this survey is on biometric characteristics that can be used for both avatar generation and ensuring security in XR.} 
% \scalebox{0.6}{
\scalebox{0.7}{
\begin{tabular}{|l|l|l|}
\hline
\textbf{Year} & \textbf{Refs}                                     & \textbf{Contribution}   \\ \hline

2016          & \cite{berntsen2016virtual}                               & Systematic review on the commercial impact of VR and the applications it is used the most.                     \\ \hline

2017          & \cite{chow2017video}                                     & Overview of cybersickness related effect on VR. Discusses how changes the attitude and behavior of person.  \\ \hline

2018          & \cite{kumari2018implementation}                          & \begin{tabular}[c]{@{}l@{}} Review the implementation issues in various VR applications. There is a special mention about the concern \\ of security the VR  systems. \end{tabular}                                                                                                                                                                                                                                             \\ \hline
2018          & \cite{bryant2020review}                                  & \begin{tabular}[c]{@{}l@{}}Reviews the papers discussing the use of VR in physical, psychological, and communication interventions. \\ Also, discusses the need  to consider the issues related to cybersecurity and cybersickness.\end{tabular}                                                                                                                                    \\ \hline
2020          & \cite{pirker2020virtual}                                 & Presents the challenges and advantages of  fully immersive VR in the computer science education.                                                                                                                                                                                                                                                                                 \\ \hline
2021          & \cite{jones2021literature}                               & Reviews the VR authentications system papers till October 2020 and highlights the research gap.                                                                                                                                                                                                                                                                                  \\ \hline
2021          & \cite{mystakidis2021deep}                                & Impact of  VR based distant learning on the privacy, security and emotions of the learners.                                                                                                                                                                                                                                                                                      \\ \hline

2021          & \cite{wen2021survey}      & \begin{tabular}[c]{@{}l@{}}
This paper reviews facial capture technologies for VR, addressing the occlusion challenges caused by headsets.\\ It identifies methodologies and research gaps, and proposes solutions. The evaluation reveals trends such as\\ marker-based tracking and  performance capture. A Modular Codec Avatar is recommended for VR facial \\capture, and an open-design headset is proposed to lower costs.\end{tabular}                                                                  \\ \hline

2022          & \cite{gumbo2022literature}                               & Analysis of privacy risks in VR and AR and potential solutions to tackle these risks.                                                                                                                                                                                                                                                                                                                                                                                             \\ \hline
2022          & \cite{giaretta2022security}      & \begin{tabular}[c]{@{}l@{}}Discuss the various attacks that can affect the confidentiality, integrity, availability, and safety of the user in \\ VR systems.\end{tabular}                                                                                                                                                                                                     \\ \hline
2022          & \cite{kurtunluouglu2022security} & \begin{tabular}[c]{@{}l@{}}Discussion of security and privacy issues in VR.  The shortcomings of information and biometric  authentication \\ systems have  been highlighted, and multi-model-based authentication has been argued as a potential solution.\end{tabular} \\ \hline

2022          & \cite{giaretta2022security}      & Discussed the various attacks that can affect the confidentiality, integrity, and safety of the user in VR systems.

\\ \hline 

2022          & \cite{kulal2022security}         & Discusses security and privacy on an integral part of metaverses like IoT, AI, Blockchain, NFT, and Crypto.                                                                                                                                    \\ \hline
2022          & \cite{odeleye2022virtually}      & \begin{tabular}[c]{@{}l@{}}Provides a taxonomical representation of various challenges in VR and further classifies these challenges into \\ the single comparative matrix.\end{tabular}                                                                  \\ \hline

2022          & \cite{dipakkumar2022systematic}                  & Discusses privacy, authorization, and data leaks in VR and AR environments.                                           \\ \hline                                                                                       2023          & \cite{huang2023security}      & \begin{tabular}[c]{@{}l@{}}This paper explores these characteristics, surveys current progress and applications, examines security and \\ privacy issues, and raises concerns about societal and human implications.\end{tabular}  

\\ \hline                                                                                       2023          & \cite{bozkir2023eye}      & \begin{tabular}[c]{@{}l@{}}This paper explores the potential of eye tracking in virtual reality (VR), highlighting its benefits for interaction \\ and understanding human cognition. It also addresses privacy concerns and proposes research directions to \\ overcome these challenges in the context of eye-tracking in VR.\end{tabular}  

\\ \hline                                                                                       2023          & \cite{qamar2023systematic}      & \begin{tabular}[c]{@{}l@{}}
This paper examines the security and privacy threats posed by immersive technologies such as extended \\ reality (XR) and metaverse. It identifies technology weaknesses, cyber security challenges, and user safety \\ concerns and proposes strategies for defense and safety measures.\end{tabular}  

\\ \hline                                                                                       2023          & \cite{heruatmadja2023biometric}      & \begin{tabular}[c]{@{}l@{}}

This study explores the challenges of security and authentication in Virtual Reality (VR) and focuses on the \\ use of  biometric authentication methods, particularly in relation to head-mounted displays (HMD). A systematic \\ literature review was conducted to investigate the commonly used biometric media, machine learning techniques,\\ and accuracy of biometric authentication in VR. \end{tabular}  

\\ \hline  

    2023          & \cite{de2019security}      & \begin{tabular}[c]{@{}l@{}}

This work gives the survey of various security and privacy aspects in Mixed reality. Risk related to investigation, \\implementation, and evaluation of data security are highlighted.  \end{tabular}  

\\ \hline  

    2024          & \cite{finnegan2024utility}      & \begin{tabular}[c]{@{}l@{}}

This scoping review summarizes the current state of behavioral biometric authentication and its potential application \\ in screen time measurement, highlighting the need for more rigorous research, particularly in child populations.
  \end{tabular}  
\\ \hline  

    2024          & \cite{kaur2024privacy}      & \begin{tabular}[c]{@{}l@{}}

This paper studies the privacy and security of data transfer from user to the Virtual Environment particularly in  \\ encryption, differential privacy and distributed learning aspect.
  \end{tabular}  
\\ \hline  

    2024          & \cite{sharma2024user}      & \begin{tabular}[c]{@{}l@{}}

A novel Metaverse Security Architecture, designed with Zero-Trust principles, is proposed, prioritizing user control\\  over data, identity, and experiences, with proactive measures ensuring a secure and immersive virtual environment.
  \end{tabular}  
\\ \hline

    2024          & \cite{amano2024visual}      & \begin{tabular}[c]{@{}l@{}}

The visual privacy issues in the metaverse are surveyed, and a method for controlling privacy is introduced, \\ enabling users to manage protection levels or delegate to an automated system.
  \end{tabular}  
\\ \hline

  2024        &  \textbf{This work}     & \begin{tabular}[c]{@{}l@{}}

%This survey paper provides the general architecture of a biometric system in extended reality and discuss all possible attacks \\ and threats. We also discuss photorealistic avatars that are integral part of XR, their generation, and possible threats in using photorealistic avatars with  biometrics characteristics in XR.
This paper presents a comprehensive overview of the fundamental structure of a biometric system within the \\ domain of extended reality,   while simultaneously addressing  various potential security vulnerabilities and threats. \\ Additionally, this paper delves into the concept   of photorealistic avatars,  which are a crucial component of XR, \\and explores the potential risks associated with incorporating  biometric features  into these virtual representations.
\end{tabular}  

\\ \hline 

\end{tabular}}
% }
\label{tab:SOTA}
\end{table*}

%There are various existing survey papers that discuss the security and privacy of the virtual reality system. In~\cite{giaretta2022security}, Alberto et.al provides the literature survey of the existing and possible threats to VR technology. They highlighted the privacy issues arising due to the combination of VR and social networks as mentioned in~\cite{o2016convergence}. They have discussed the various attacks that can affect the confidentiality, integrity, availability, and safety of the user in VR systems. Further, the existing biometric systems for VR  are discussed and their performances are listed. This survey paper provides the general architecture of the biometric system in virtual reality (refer to figure~\ref{VR_workflow_attacks}) and shows all the possible attacks and threats at each and every step. We further discuss the countermeasures to be taken to protect the VR from these attacks.

% are discussed and possible vulnerabilities are identified.
%The chronological order of the paper is done according to the figure~\ref{VR_organisation}
%\begin{figure*}[]
%\includegraphics[height= 150pt,width= 150pt    ]{Figures/OrganizationOfPaper.png}
%\caption{Organisation of paper.}
%\label{VR_organisation}
%\end{figure*}

\textbf{Methodology:} The rapid evolution of XR applications has resulted in the development of several technologies, and research has focused on addressing security and privacy issues. Figure~\ref{fig:VR_countVsYear} shows the number of research papers published on security and privacy in XR applications, indicating an upward trend in this topic. 

We approach the research survey by formualating and addressing the following research questions (RQ):
\begin{itemize}
    \item \textbf{RQ1:} What are the major vulnerability points and possible attacks in the biometric based XR system workflow?    
    \item \textbf{RQ2:} What are the existing biometrics characteristics that are employed in XR and are they vulnerable for attacks?
    \item \textbf{RQ3:}  How has avatars evolved over the years to give photo-realistic quality with biometric characterics? What are the Deep learning techniques used for avatar generation?
    \item \textbf{RQ4:} What are the currently available datasets for the biometrics of XR and avatar generation and verification?
    
\end{itemize}

In the survey we collected $318$ articles from Google scholar. The steps followed in filtering and selection of the articles are:

\begin{itemize}
    \item \textit{Searching the database:}  The articles consists related to any form of biometrics used in the XR domain, attacks in the XR work flow, avatar generation and verification, deep learning models for avatar generation and the dataset used across the authentication in XR. The keywords used are (1) "Security" and "privacy" in "Virtual/Augmented Reality", (2) Security "Attacks" in "Virtual/Augmented Reality", (3) "Avatar generation" for "Virtual/Augmented Reality", (4) "Biometrics" in "VR/AR" (5) "Spoofing attacks" on "VR/AR".
    \item \textit{First screening:} Read through the title, abstract, conclusion and methodology used in the searched papers to find the relevant papers in the scope of this survey.
    \item \textit{Second screening:} In the second phase, we conducted a comprehensive review of the 62 shortlisted papers out of the initial 391 articles. These 62 articles represent a refined sample space, offering a clear understanding of the various components of XR. The review of these 62 papers is systematically presented across the different sections of this survey.
    \item \textit{Analysis step:} We finally provide the insights from the review.
\end{itemize}

Furthermore, the popularity of security and privacy in XR has resulted in several survey papers focusing on different aspects of security and privacy. Table~\ref{tab:SOTA} presents the contributions of the existing survey papers that address security and privacy in XR applications. Existing survey papers discuss the general framework for privacy and security tailored to different applications, including blockchains, the IoT, and social networks. Biometrics has become an integrated part of XR systems to authenticate the user and generate avatars to represent the user in the virtual space. There is no exclusive survey that can provide a good overview of XR biometrics.  Although the utility of biometrics is discussed in \cite{kurtunluouglu2022security}, a detailed discussion of biometric utility for XR applications has not been presented. Therefore, in this paper, we present a comprehensive review of biometric solutions for XR applications. In particular, we discuss the vulnerability of the XR system and existing biometric solutions studied in the literature for XR applications. To the best of our knowledge, this is the first study to discuss the potential of biometrics in XR applications. The contributions of this study are as follows:

\begin{itemize}

 \item 	Comprehensive overview of existing biometric  solutions that can be used for authenticating user and also to generate the avatar representing the user in  XR environments. 
 \item 	Present and discuss different vulnerability points and possible types of attacks on general XR systems. This is the first study to present a systematic representation of the vulnerabilities of XR systems. 
 \item 	Extensive discussion of both physiological and behavioral biometrics for XR system authentication. 
 \item 	 A comprehensive overview of biometric-based photo-realistic avatar generation techniques that are widely used in XR applications.
 \item 	Extensive discussion of future work indicating the potential of biometrics in XR applications.
   
\end{itemize}

\begin{figure}[]
\centering
\includegraphics[height= 300pt, width= 245pt    ]{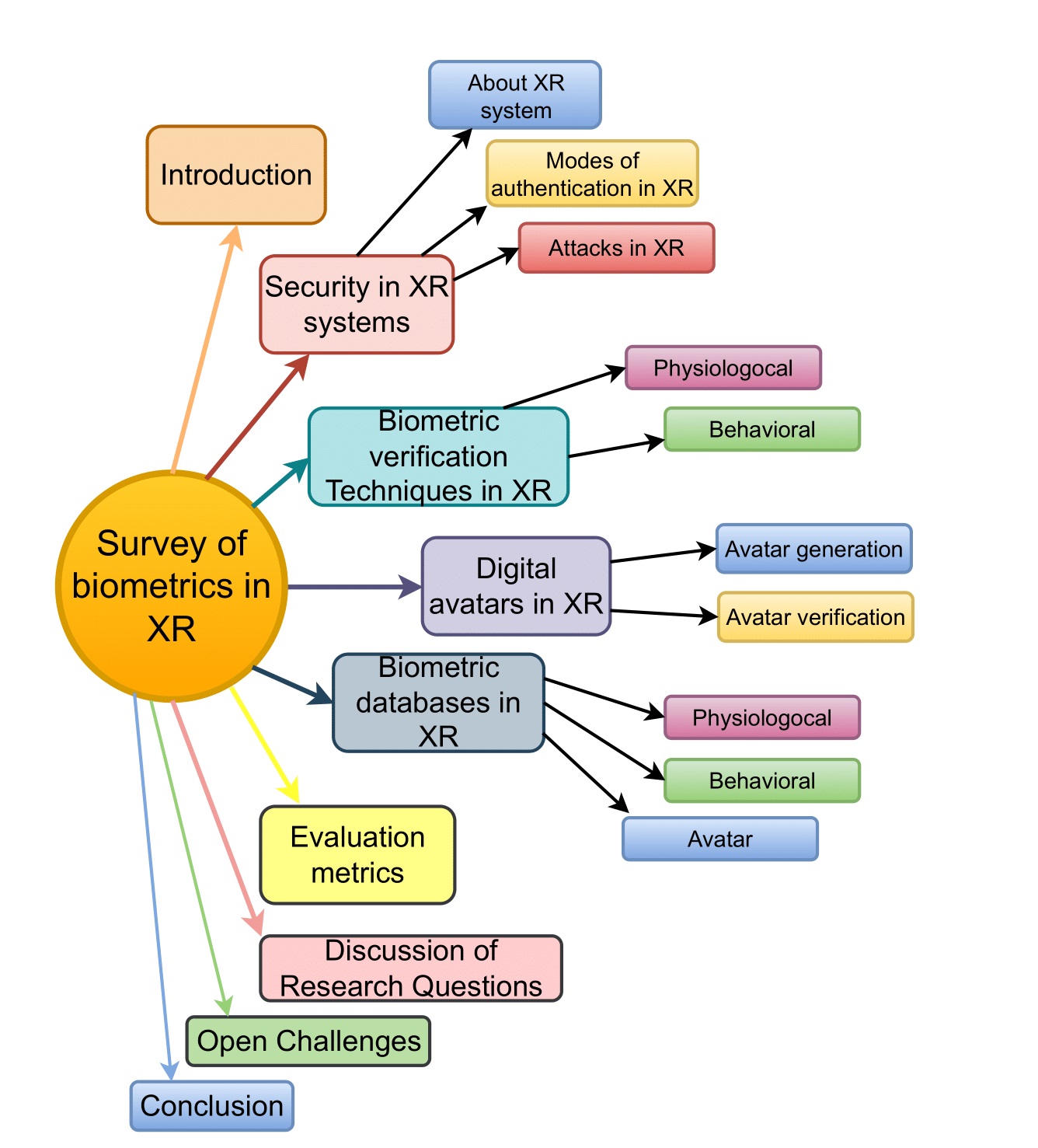}
\caption{Taxonomical representation on the structure of this paper}
\label{TaxonomySurvey}
\end{figure}

The organization of the paper follows the taxonomy mentioned in Figure~\ref{TaxonomySurvey}. Section~\ref{Security in XR systems} describes the current workflow of the XR system, the location of vulnerability, and possible attacks at these vulnerability points. We also discuss various authentication schemes used for the authentication of the XR system. In Section~\ref{BiometricVR} we discuss the various physiological and behavioral modes of biometric verification techniques for XR. Section~\ref{Avatars} discusses the avatar generation and verification methodologies. In Sections ~\ref{Databases} and ~\ref{PerformanceMetrics} we highlight the existing databases and performance metrics, respectively. Section \ref{Sec:QA} discuss the research questions and Open challenges in current XR research are discussed in section ~\ref{OpenChallenges}. Finally, conclusion is discussed in the Section~\ref{Conclusion}.

\section{Security in XR systems}
\label{Security in XR systems}
In this section, we discuss the security aspects of XR systems, focusing on the different vulnerability points that can allow attackers to perform various types of attacks on XR systems. First, we present an overview of the XR system with the main working blocks and discuss different types of authentication mechanisms. We then discuss different types of attacks that can be performed on the different vulnerable parts of XR systems.  

\subsection{XR System}
%\textit{- Discuss on working principle of VR with the aid of figures (illustrate individual operation with figures)
%}

\begin{figure}[]
\centering
\includegraphics[height= 225pt,width= 350pt    ]{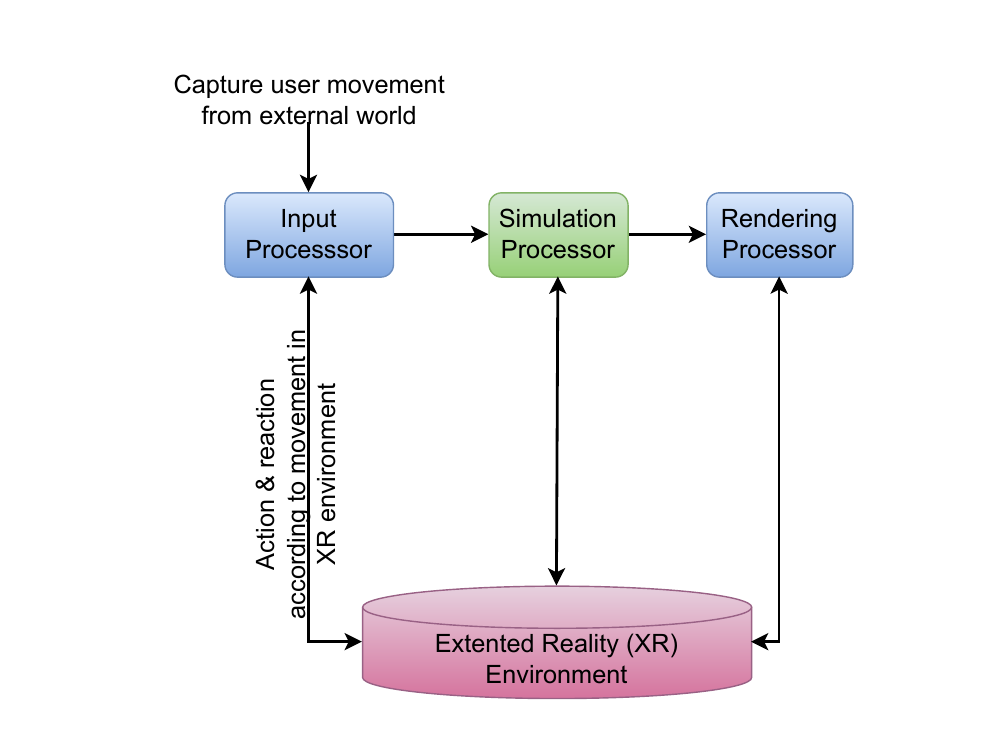}
\caption{Illustrating the working principle of extended reality}
\label{VR_principle}
\end{figure}

Extended reality (XR) systems are composed of four key components that collaborate harmoniously to craft an immersive experience for the user, as depicted in Figure~\ref{VR_principle}. These components encompass the input processor, simulation processor, rendering processor, and the XR environment. The input processor captures and processes the user's movements and interactions within the XR environment, thereby enabling seamless interaction with the extended world in an intuitive manner.
The simulation processor is responsible for fabricating an extended reality environment and dictating its responses to the user input. This entails simulating the orientation and locomotion of extended objects, along with other tangible phenomena, such as physics and illumination. This helps create a real and believable XR world that responds accurately to what users do. The simulation processor ensures that the XR environment behaves consistently and predictably. This provided users with a smooth and exciting experience.

The rendering processor creates sights and sounds that make up what a person experiences through their senses, such as seeing things in 3D, hearing sounds, and feeling touch-like sensations. This is important because it helps make an experience feel real, as if a person is actually inside a different world. The rendering processor works closely with another part called the simulation processor, which replicates the live experience. This makes the entire experience feel accurate and responsive to what a person is doing.

The XR environment is like a special place where all instructions (scripts) and things (entities) exist in the extended world. It also includes rules that control how these things operate. This special place defines how the extended world looks and how things act in it. It also demonstrates how different parts of the extended world work together. The XR environment is important because it helps to build a connected and believable extended world that makes sense and feels real to the person using it. In summary, the extended reality system has four main parts that work together smoothly to provide the user with a continuous and deeply engaging experience. It does this by taking into account what the user does, creating very realistic computer-made scenarios, showing things in both sight and sound, and setting up a consistent alternate world. This system is good at taking the user to another world where they can interact in a natural and easy manner.

\subsection{Various modes of authentication in XR}
\label{ModesOfAuthentication}

\begin{figure*}[]
\centering
\includegraphics[height= 250pt,width= 325pt    ]{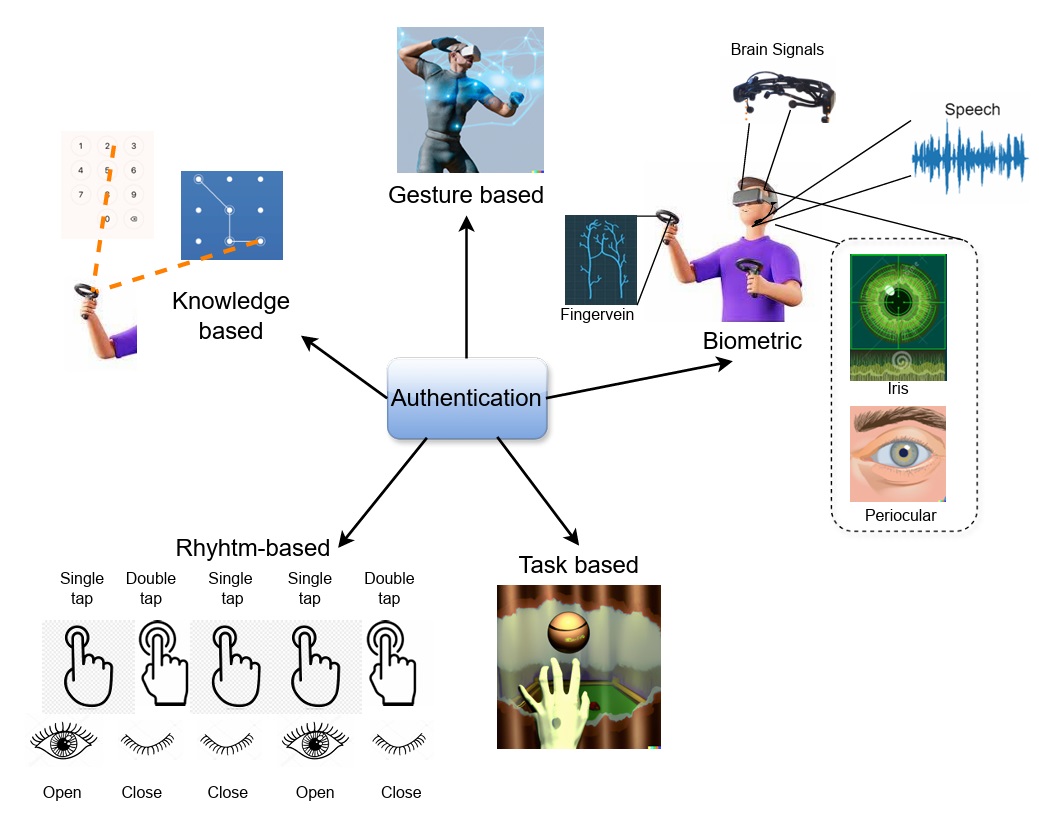}
\caption{Types of authentication widely used in XR scenario that includes both knowledge based and biometrics.}
\label{VR_biometric}
\end{figure*}

Various modes of authentication have been developed over the years to protect XR users from attacks. These authentications are broadly classified as knowledge-based, gesture-based, gaze/task-based, or rhythm-based. Figure~\ref{VR_biometric} shows the different types of user authentication that are widely employed in  existing XR systems. 

\begin{itemize}

\item \textbf{Knowledge-based authentication} refers to the traditional method of authentication where users have to enter a personal identification number (PIN) or password to unlock a device or access a system [\cite{george2017seamless, olade2020exploring}]. Researchers have explored various transformations and innovations to enhance the security of this type of authentication.  One approach is the use of shuffled virtual keyboards, which randomize the layout of the keys to prevent shoulder surfing attacks and password guessing. Rajarajan et al. proposed a shoulder-surfing-resistant authentication system that employs a shuffled virtual keyboard, making it challenging for an attacker to observe the user's keystrokes [\cite{rajarajan2014shoulder}]. Similarly,~\cite{holland2018dynamic} introduced a dynamic virtual keyboard that continuously changed its layout, further enhancing security against shoulder surfing attacks. Another interesting transformation is RubikAuth, proposed by~\cite{mathis2020rubikauth} which integrates the concept of a Rubik's cube into the authentication process. Users manipulate the virtual Rubik's Cube to generate a password by adding an additional layer of complexity to the authentication process. RubikBiom, also proposed by Mathis et al., combines knowledge-based authentication with biometrics by using the orientation of the Rubik's Cube as a biometric feature [\cite{mathis2020knowledge}]. The fusion of knowledge-based and biometric authentication aims to enhance both security and usability.       

\item \textbf{Gesture-based authentication} refers to a method of user authentication that utilizes specific hand or body movements to verify the identity of an individual. Instead of relying solely on traditional methods, such as passwords or PINs, gesture-based authentication leverages the unique patterns and characteristics of user gestures to grant access to systems or devices. However, such actions can be cumbersome and awkward, particularly when performed in public places. Future research on gesture-based biometric authentication should focus on developing techniques that capture distinctive biometric features in a non-intrusive and natural manner. This ensures that the authentication process is user-friendly and does not draw unnecessary attention, thereby promoting both usability and security in various contexts.

\item \textbf{Task-based authentication} is a novel approach that leverages the innate consistencies in human performance of everyday tasks to establish a unique signature for individuals in extended reality (XR) environments. Humans develop a certain level of proficiency and consistency in performing common tasks owing to their real-world experience. Activities such as throwing a ball, lifting a chair, swinging a golf club, or driving a car involve specific patterns of movement that are unique to each individual. In XR environments, task-based movements can be captured and analyzed to create a personalized authentication system. By recording and analyzing the performance of these tasks, a unique signature was established based on individual movement patterns, timing, and coordination. However, challenges remain in the implementation of task-based authentication systems. This requires the accurate tracking of user movements using sensors or cameras within the XR environment. The system must also be able to distinguish between genuine users and imposters attempting to mimic the task movements.

\item \textbf{Rhythm-based authentication} is a unique approach where users respond to a predefined rhythm or pattern during the enrollment process. This authentication method relies on the ability of the user to accurately reproduce rhythm to verify their identity. The response can be based on various actions such as tapping a finger, blinking an eye, or any other predefined gesture. During enrollment, users were prompted to synchronize their actions with a specific rhythm or sequence. The system captures and records the user response as a unique rhythm-based signature. When authentication is required, the user is asked to reproduce the same rhythm or pattern with which they initially enrolled. The system compares the user's response to the enrolled rhythm and determines whether it matches, grants, or denies access accordingly.

\item \textbf{Biometric:} Biometric characteristics are widely employed in the XR for authentication that includes:
    \begin{itemize}
%    \item \textit{Fingerprint-based biometrics} have shown promising reliability in terms of security. Fingerprint-based authentication is mostly performed on the hand controllers of VR devices.
    \item \textit{Iris:} The use of various functionalities of the iris as the biometric to detect the authentication of the user. The advantage of Iris-based biometrics is that they are not prone to shoulder surfing attacks.~\cite{zhu2020blinkey} proposed a two-factor authenticator called BlinKey. The first factor is blinking of the eye in a certain rhythm. The second factor is the variation in the pupil size at each blink. These blinks and size variations were captured by the eye tracker. BlinKey-based authentication relies on a blinking pattern, so an HMD-based solution that captures the iris and a partial periocular region is used for biometric authentication [\cite{boutros2020iris}]. Features for iris extraction for this purpose include ordinal measures (OM) [\cite{sun2008ordinal}] and Discrete Fourier Transforms (DFT) [\cite{miyazawa2008effective}]. Ordinal Measures focus on the relative ordering of pixel intensities, making them robust to lighting variations, noise, and contrast changes in iris recognition. OM encodes these relationships into ordinal codes for feature extraction. DFT capture the frequency components of the iris texture, highlighting periodic patterns that aid in distinguishing irises despite variations in scale and orientation. In addition to this Convolution Neural Network (CNN) based deep learning feature extraction methods like DeepIrisNet [\cite{gangwar2016deepirisnet}] and MobileNetV3 [\cite{howard2019searching}] is utilized in 
 [\cite{boutros2020fusing}] for highly accurate iris recognition.

    % \item Speech: 
% Electroencephalography (Electroencephalography (
\item \textit{Brain signal-based biometrics}, specifically the use of electroencephalogram (EEG) signals, has gained interest as a non-intrusive input modality, particularly in conjunction with wearable headsets like virtual reality (VR) devices~\cite{hertweck2019brain}. EEG signals were collected using electrodes placed on the scalp to measure the electrical activity of the brain. One challenge in utilizing EEG data for Brain-Computer Interface (BCI) algorithms is the poor generalization performance across users. EEG signals exhibit inter-user differences owing to variations in the brain structure and activity patterns. However, these differences can be leveraged for user authentication. By studying and understanding the inter-user differences in EEG signals, researchers aim to develop authentication systems based on these unique brain activity patterns. Each individual's brain signal can serve as a distinctive biometric [\cite{li2019brain}]. This approach offers a non-intrusive and potentially more secure authentication method. In EEG signal analysis, feature extraction methods like the Wavelet Packet Transform (WPT) [\cite{bong2017implementation}] and Autoregressive (AR) models [\cite{makhoul1975linear, agarwal2020vop}] are crucial. WPT offers detailed decomposition, capturing both time and frequency information, while AR models can predict future signal values, formulated either as a spectral estimation problem or a linear prediction in the time domain. Power Spectral Density (PSD) analysis, computed via the Fast Fourier Transform (FFT) and refined using Welch’s method, characterizes signal power distribution over frequency. Additionally, Statistical Parameters of Signals (SPS) measure the distribution and complexity of EEG signals, extending concepts like cumulants and entropy is also used as feature of EEG signal. These statistical features, while useful for providing insights into its overall distribution, are generally not well-suited for capturing complex, time-varying components of signals like those found in EEG. EEG signals are inherently represented as time-series data, necessitating the extraction of nuanced temporal dependencies that characterize these signals. Gated Recurrent Units (GRUs) are particularly well-suited for this task, demonstrating the ability to capture and model the complex temporal relationships embedded within EEG data. By leveraging GRUs, the temporal dynamics of EEG signals can be effectively unraveled, allowing for a more comprehensive understanding of the underlying neural processes [\cite{luo2024feature}].

    \end{itemize}  
    \item \textbf{Continuous authentication:} Continuous authentication in Extended Reality (XR) refers to the ongoing process of verifying a user's identity and intent while they interact within virtual, augmented, or mixed reality environments. Unlike traditional authentication methods that occur only at login, continuous authentication employs real-time biometric and behavioral data analysis. This includes factors such as head movement patterns, gaze direction, and face and voice characteristics. By assessing these factors, XR systems can detect anomalies that might indicate unauthorized access or compromised interactions. This dynamic approach enhances security by adapting to user behavior changes and prevents unauthorized users from gaining extended access to sensitive XR applications and data. Continuous authentication thus safeguards user experiences and privacy, maintaining a secure environment throughout the user's XR session.
\end{itemize}

\subsection{Various attacks in an extended reality}

\begin{figure*}[h]
\centering
\includegraphics[height= 325pt,width= 425pt]{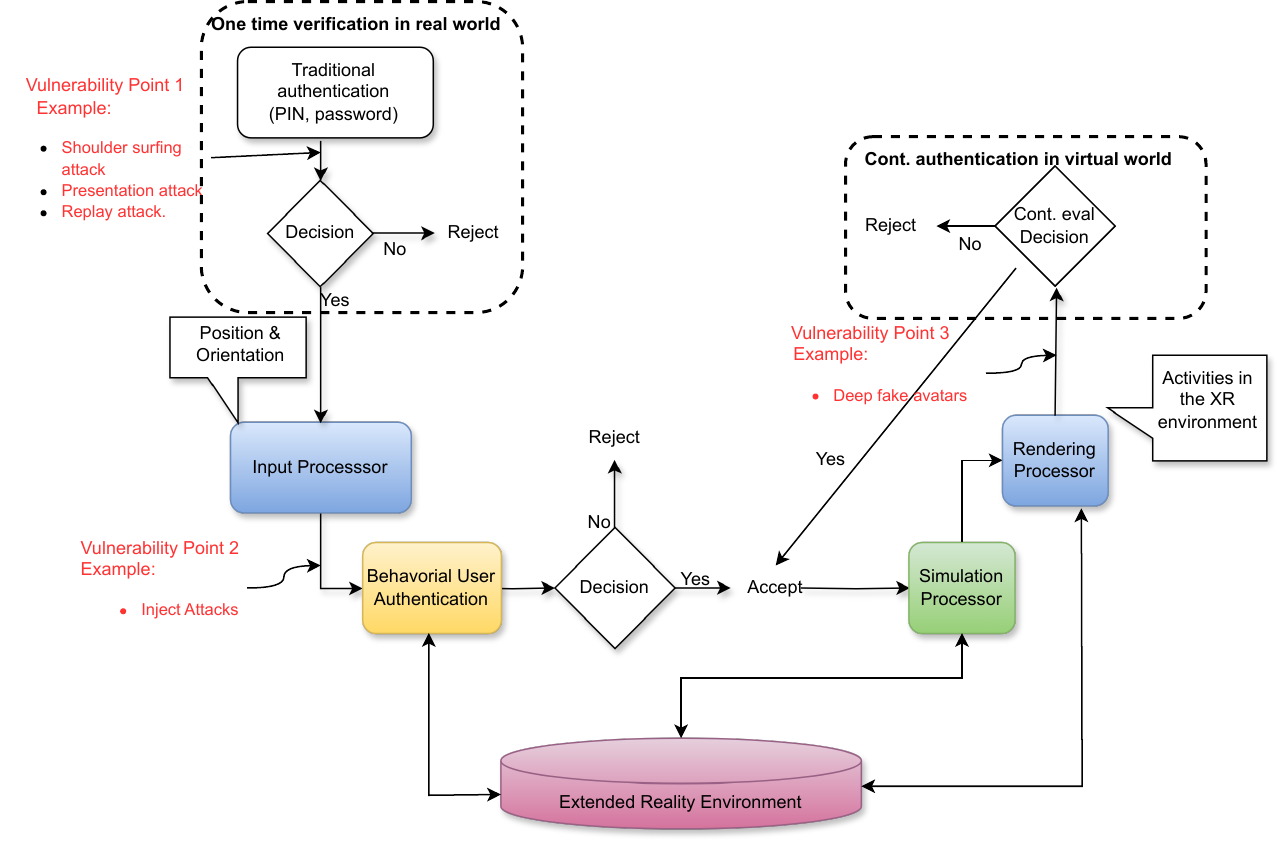}
\caption{Illustration of various vulnerability   of XR system in which the biometric characterics (physiological and behavioral) are involved. In total, three vulnerability points need to be addressed to achieve a secure and trustworthy XR system. }
\label{fig:VR_workflow_attacks}
\end{figure*}

Figure~\ref{fig:VR_workflow_attacks} illustrates the vulnerabilities and potential points of attack across different levels of an XR system. The three primary levels, where user authentication can be performed, are   visually represented within a dashed rectangular box in Figure~\ref{fig:VR_workflow_attacks}. The first level of authentication can be performed on the device before using the device for XR applications. Second-level authentication is based on continuous authentication using gestures and actions that can be performed during interactions in an XR world. Third-level authentication is performed in virtual space rendering, where the authentication of rendered avatars in the XR space is  carried out. Therefore, depending on the various functional blocks of the XR system, we identified four different vulnerability points, as illustrated in Figure~\ref{fig:VR_workflow_attacks}. In the following, we discuss different types of attacks that can be foreseen at different vulnerability points of XR systems. 

\subsubsection{\textbf{Vulnerability Point 1}}
Vulnerability point 1 is mainly associated with early authentication, primarily performed to validate the XR devices to authenticate the legitimate user. Conventional authentication methods include PINs, cards, passwords, and biometrics.  Possible sources of attacks include direct attacks on biometrics and   shoulder-surfing attacks on passwords and PINs. 

The \textit{Direct attacks or presentation attacks} [\cite{10.1145/3038924}] on the XR devices include the presentation of the biometric artefacts of the legitimate user to gain malicious access. The attacker can generate biometric artifacts either by printing or by synthesizing and presenting them directly to the biometric sensors. Depending on the type of biometric characteristics, the attack artifacts and generation of Presentation Attack Instruments (PAI) for successful attacks also vary. For instance, print attacks involve using printed biometric data (such as a fingerprint or face image) to deceive the system. In 3D mask attacks, a three-dimensional mask resembling the legitimate user's face is created and presented to the sensor. Replay image, speech or video attacks involve playing back a recorded image, speech or video of the legitimate user to fool the device. For details on presentation attacks on biometric systems, readers can refer to~\cite{10.1145/3038924, marcel2019handbook}. Several defense mechanisms have been proposed for PAI attacks. Optical flow texture analysis, for instance, detects subtle motion inconsistencies in spoofed biometric data by tracking pixel-level movements, effectively distinguishing real subjects from static or fake presentation attacks [\cite{li2022face, bhattacharjee2019recent, damer2016practical, raghavendra2015presentation, ramachandra2017presentation}]. Attacks on speech involve generating deepfakes or replaying audio to deceive authentication systems. These attacks exploit weaknesses in speech features like pitch, formants, and phonemes [\cite{kumar2022fake, sadashiv2023source}]. Deep learning models, such as ResNet [\cite{magazine2022fake}] and ASSIST are employed to analyze these features, detecting subtle inconsistencies in the generated audio that distinguish authentic speech from manipulated or replayed versions. 

  \textit{Shoulder surfing attacks} in XR are similar to their counterparts in the physical world. They involve an attacker trying to obtain sensitive information by watching the actions and movements of the target user. In the XR context, an attacker may be able to see the user's virtual keyboard or other input devices and use that information to steal passwords or other sensitive data. In addition, the attacker may be able to observe the user's virtual interactions and use that information to gain unauthorized access to restricted areas or sensitive information. These attacks can be especially dangerous in virtual environments that are used for sensitive tasks such as military training or healthcare.

\subsubsection{\textbf{Vulnerability Point 2}}

 Vulnerability point 2 is mainly targeted at the communication channel that connects the user interaction to the XR environment. User interaction is captured in terms of head movement, hand movement, full-body movement, and user behavior based on the stimuli. The possible sources of attacks include the injection of human behavior through a side-channel attack, as detailed below.

\textit{Side channel attacks} could be used to extract sensitive information from the user, such as biometric data or other personal information. For example, an attacker could use a microphone placed near the user's HMD to capture the sounds of their hand movements and use that information to reconstruct their hand gestures and steal biometric data [\cite{al2021vr}]. Another example is the use of sensors to capture a user's body movements and infer sensitive information about their physical health or other personal characteristics. Additionally, inject attacks [\cite{pant2024mixed}] could be performed by overriding the biometric capture device. In such scenarios, an attacker could inject a biometric sample corresponding to the victim directly into the device, bypassing the actual user interaction. This could allow the attacker to impersonate the victim within the XR environment, gaining unauthorized access to sensitive areas or information. These injected biometric samples could be sourced from previously captured data or generated using advanced techniques like deepfakes, making the attack difficult to detect. 

To counter inject attacks in Extended Reality (XR) environments, effective defense strategies include implementing tamper-resistant biometric capture devices to prevent unauthorized access and utilizing robust encryption protocols for secure data transmission. Incorporating multi-factor authentication methods, such as PINs or behavioral biometrics, enhances security. Additionally, deep fake detection algorithms can monitor user behavior and identify unusual patterns indicative of potential attacks, strengthening overall system integrity [\cite{qamar2023systematic}].

\subsubsection{\textbf{Vulnerability Point 3}}      
Vulnerbality point 3 is the target on the rendered virtual space where the attacks can impersonate the virtually rendered avatar. Advanced XR systems continuously authenticate users in virtual space, especially in confidential meetings [\cite{sivasamy2020vrcauth}]. However, secure authentication is assured based on biometric characteristics, including avatar verifications, voice authentication [\cite{desplanques2020ecapa, agarwal2022significance}], and gaze movement [\cite{yang2023secure, turkmen2023put}].  In the following section, we discuss the possible attacks at vulnerability point 4.

\textit{Man-In-The-Room attack} is a type of attack where the attacker joins the target's VR environment without their knowledge and remains invisible while extracting information or manipulating the environment. This attack is particularly concerning in VR environments because users tend to have a high level of immersion and may not notice the presence of an attacker. An attacker can use this attack to spy on the user's interactions, collect sensitive information, or even manipulate the environment in a way that could be harmful to the user. For example, by leveraging morphing techniques, the attacker can subtly alter avatars or objects within the environment, making changes that go unnoticed by the user but that could influence their behavior or decisions. Additionally, the use of synthetically generated images or videos can further enhance the attack, allowing the intruder to introduce realistic yet fake content, such as deepfake avatars or manipulated scenes, which can deceive or mislead the user. These manipulations can be carried out by exploiting vulnerabilities in the VR system or by using social engineering techniques to trick the user into granting access to the VR environment. As VR has become more widespread and integrated into various industries, it is important to address the potential security threats posed by man-in-the-room attacks, including the sophisticated use of morphing and synthetic data generation, to ensure the safety and privacy of VR users.

Looking into the adverse affects that these fake avatars create,~\cite{bader2014securing} proposes a watermarking algorithm for securing avatar access in virtual worlds by embedding a 128-bit biometric fingerprint into the ordered facet-vertex rings of the avatar's face, ensuring imperceptibility and resistance to potential attacks during authentication [\cite{bader2016identity, lin2022digital}].

\textit{Camera Stream and Tracking Exfiltration attack} is a type of attack where the attacker gains unauthorized access to the head-mounted display's (HMD) live stream and the HMD front-facing camera stream. By gaining access to these streams, attackers can potentially monitor the user's actions and surroundings, compromising their privacy and security. In addition, the attacker can track the user's head movements and use this information to gain a better understanding of their environment. This type of attack can be particularly concerning in situations where the user is engaging in sensitive activities such as financial transactions or national security discussions. Therefore, it is essential to implement robust security measures to prevent unauthorized access to HMD streams.

\section{Biometric verification Techniques in XR}
\label{BiometricVR}
In this section, we discuss the existing biometric verification techniques that are widely adopted in XR applications. Biometric characteristics can be classified into two categories: physiological and behavioral. Physiological biometrics involves the analysis of physical measurements of the human body. Examples of physiological biometrics include iris-recognition and fingerveins. On the other hand, behavioral biometrics analyzes body movements such as voice, gait,  and keystrokes.
% ~\ref{}. 

\begin{figure}[htp]
\centering
\includegraphics[height= 300pt,width= 250pt]{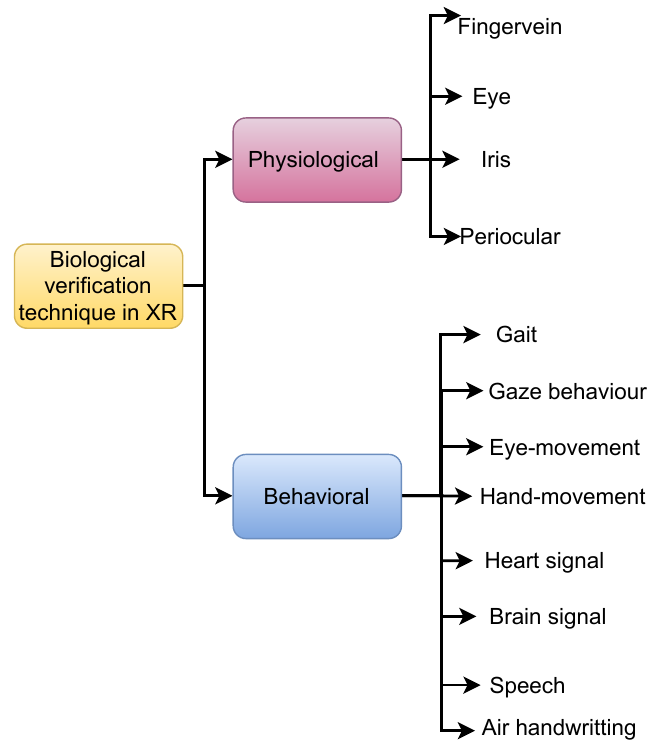}
\caption{The taxonomic representation of biometric verification techniques in XR illustrates the applicability of both physiological and behavioral biometrics.}
\label{fig:TaxonomySection3}
\end{figure}
 
\subsection{Physiological biometric}

%%%%%%%%

\begin{table*}[h]
\centering
\caption{Summary of existing studies on physiological biometric characteristics employed in  Extended Reality (XR) systems.}
\scalebox{0.65}{
\begin{tabular}{|c|c|c|c
|l|}
\hline
\textbf{Author}   & \textbf{Year} & \textbf{Biometric Modality}   & \textbf{Device}   & \multicolumn{1}{c|}{\textbf{Description}}                                                                                                                                                                                                                                     \\ \hline
~\cite{tran2023anti}     & 2023          & Fingerveins        &  HTC - Vive        & \begin{tabular}[c]{@{}l@{}}Fingerprints and palms have the risk of presentation attacks. In this work, fingerveins \\ are used as VR biometrics. Fingerveins are secure and difficult to imitate as they \\ lie beneath the skin\end{tabular}                                  \\ \hline
~\cite{luo2020oculock}  & 2020          & Eye lid+Extraocular & Designed Smart Glass   & \begin{tabular}[c]{@{}l@{}}The use of eye gaze has low stability and high error. In this work, the use of the eyelid, \\ extraocular, and surrounding nerves instead of just the eyelid is proposed. It is shown \\ that the multi-model approach is more reliable\end{tabular} \\ \hline
~\cite{boutros2020iris}   & 2020          & Iris+Periocular &  Head mounted display    & \begin{tabular}[c]{@{}l@{}}Iris and periocular regions when used independently have lower authentication accuracy. In \\ In this work, the score-level fusion of the iris and periocular is done to improve accuracy.\end{tabular}                                            \\ \hline
~\cite{zhu2020blinkey}       & 2020          & Iris & Microsoft Hololens 2              & \begin{tabular}[c]{@{}l@{}}Two-factor authentication is proposed in which along with the user's iris data a secret \\ password is entered in the Blink of an eye (BlinKey). The rhythms in the BlinKey are \\ captured with an eye tracker in the headset.\end{tabular}          \\ \hline
\end{tabular}}
\label{Physio}
\end{table*}

Physiological biometrics has been widely developed to secure XR systems. Table~\ref{Physio}  summarizes the use of different physiological biometric characteristics in XR systems. Fingerveins are located beneath the skin, which makes them more secure and adaptable [\cite{tran2023anti}]. Because of their subcutaneous position, they are difficult to imitate and pose a lower risk of presentation/spoof attacks. Therefore, the adoption of fingerveins as a biometric measure in XR mitigates the limitations associated with traditional authentication methods, offering a promising solution for secure and reliable identity verification in the metaverse.

To enhance biometric authentication in Extended Reality (XR) systems, iris data-based biometrics have been integrated into the head-mounted display. However, this approach is not without its limitations. One challenge is the potential distortion of the iris data, which can result in misleading responses. To address this issue, a data-augmentation technique has been proposed by~\cite{varkarakis2018deep}. By augmenting iris data with off-axis images, as well as images with blur, contrast, and shadow, the accuracy and robustness of iris authentication can be improved. Another concern with iris-based biometrics is their vulnerability to presentation attacks owing to the static nature of iris authentication. To mitigate this risk, a real-time iris detection method was introduced by ~\cite{adnan2022robust}. This method aims to detect and prevent presentation attacks by continuously monitoring the iris in real-time during the authentication process. By dynamically analyzing iris features, the system can detect anomalies and suspicious behaviors, enhancing the security of iris-based biometric authentication.

In addition to iris-based biometrics, the periocular region has been recognized as a valuable source of unique user information [\cite{boutros2020iris}]. The periocular region refers to the area surrounding the eye, including the eyelid, extraocular cells, muscles, and the surrounding nerves. By fusing the iris and periocular regions at the score level, researchers have demonstrated enhanced performance in biometric authentication [\cite{boutros2020iris}]. This fusion approach leverages the complementary information provided by the iris and periocular regions, thereby improving the overall accuracy and reliability of the authentication system.

\begin{table*}[htp]
\centering
\caption{Summary of existing studies on behavioral biometrics employed in Extended Reality (XR) systems.}
\scalebox{0.6}{
\begin{tabular}{|c|c|c|c|l|}
\hline
\textbf{Author}            & \textbf{Year} & \textbf{Biometric Modality}    & \textbf{Device}                                                                          & \multicolumn{1}{c|}{\textbf{Description}}                                                                                                                                                                 
\\ \hline
~\cite{winkler2022questsim}           & 2022          & Gait  & Headset and Controller                                                                                 & \begin{tabular}[c]{@{}l@{}}This study introduces a reinforcement learning framework designed to produce lifelike full-body \\ movements using limited input signals received from wearable devices. Even in situations where \\ data regarding the lower body is missing, the framework successfully replicates leg movements \\ that closely match the actual movements, and it can adjust to various walking styles, body \\ dimensions,  and settings. This adaptability makes it well-suited for applications in augmented \\ reality (AR)  and virtual reality (VR).\end{tabular}                                                                  

\\ \hline
~\cite{miller2022behavior}           & 2022          & Eye Movement & -                                                                                  & \begin{tabular}[c]{@{}l@{}}A novel method involves utilizing eye movement patterns as behavioral biometrics for conducting \\ robust user authentication across various systems within the virtual reality (VR) environment.\end{tabular}                                                                                                                                                                                                                                                                \\ \hline
~\cite{lohr2022demonstrating}             & 2022          & Eye Movement  & Vive Pro Eye                                                                                & \begin{tabular}[c]{@{}l@{}}The potential exists for the motion behavior of users in virtual reality (VR) to serve as a unique \\ biometric signature, allowing for continuous user identification and authentication. This method \\ can be used without jeopardizing the security of the VR application, even if traditional passwords \\ are obtained by malicious individuals.\end{tabular}                                                                                                                                                           \\ \hline
~\cite{wierzbowski2022behavioural} & 2022          & \begin{tabular}[c]{@{}c@{}} Eye-Tracking and \\  VR Headset \\ Orientation\end{tabular} & HTC Vive

&\begin{tabular}[c]{@{}l@{}}This work centers on identifying users through the combination of eye-tracking and VR headset \\ orientation with 360-degree VR videos that have not been previously seen. During the enrollment \\ process, users are presented with a 360-degree video. In the authentication phase, they encounter \\ an unfamiliar 360-degree video. The primary objective is to recognize and verify users based on \\ their natural and unpressured behavioral responses.\end{tabular}                                                                                  \\ \hline
~\cite{virtualworlds1010004}           & 2022          & Eye Gaze & HTC Vive Eye Pro                                                                            & \begin{tabular}[c]{@{}l@{}}
Compared to previous research that necessitated extensive data from a wide range of sensors, \\ this study offers a minimal eye gaze feature for user identification while maintaining accuracy \\ levels similar to those achieved in earlier works.\end{tabular}                                                                                                                                                                                                         \\ \hline
~\cite{liebers2022identifying}           & 2022          & Hand Tracking  &   Microsoft HoloLens 2                                                                             & \begin{tabular}[c]{@{}l@{}}Using passwords carries inherent risks. Therefore, hand-tracking biometrics have been introduced \\ to enhance security. This technology identifies users based on their unique finger movements, both \\ uni- and bimanual, which are collected during their interactions with eight diverse universal interface \\ elements, including buttons and sliders.\end{tabular}                                                                                                                       \\ \hline

~\cite{sun2022perae}                          & 2022                               & Heart Signals       &  -        & \begin{tabular}[c]{@{}l@{}}The proposal suggests Electrocardiogram (ECG)-based Identity Recognition (EIR), claiming that \\ it is less susceptible to vulnerabilities due to its reliance on internal biological features.\end{tabular}                                                                                                                                                                                                                                                                                 \\ \hline

~\cite{liebers2021using}           & 2021          & Gaze Behaviour  & Oculus Quest       &                                                                     \begin{tabular}[c]{@{}l@{}}A biometric identification system utilizes an individual's unique gaze behavior. More specifically, \\ the system places emphasis on the user's gaze patterns and head orientation when tracking a \\moving stimulus.\end{tabular}                                                                                                                                                                                  \\ \hline
~\cite{miller2022temporal}   & 2021          & Motion & Oculus Quest                                                                    & \begin{tabular}[c]{@{}l@{}}
User identification can encounter issues when a user employs one device for enrolling their motion \\ trajectory and another for authentication. This study utilizes a siamese neural network to determine \\ the disparity between the enrolled and tested trajectories from different devices. The closest distance \\ found is considered the identified trajectory.\end{tabular}                                                                                    \\ \hline
~\cite{olade2020biomove}             & 2020          & Kinesiological  & Oculus Quest                                                              & \begin{tabular}[c]{@{}l@{}}Each person possesses unique kinesiological traits, encompassing their distinctive behavioral and \\ movement characteristics. These traits can be harnessed in security-sensitive virtual reality (VR) \\ applications to offset users' incapacity to identify potential observers in the real world.\end{tabular}                                                                                                                                      \\ \hline
~\cite{pfeuffer2019behavioural}         & 2019          & Body Motion     & Oculus Quest                                                                           & \begin{tabular}[c]{@{}l@{}}This work examines body movements as a form of behavioral biometrics for virtual reality.\\ Specifically, they are exploring which behaviors can effectively be used to distinguish and identify \\ a user.\end{tabular}                                                                                                                                                                                                                                                         \\ \hline

~\cite{li2019brain}                      & 2019                               & Brain Signals & BCI Headset                          & \begin{tabular}[c]{@{}l@{}}In this study, EEG data collected from users in both VR (individuals using VR headsets)\\ and non-VR  (those utilizing laptops) scenarios is examined and assessed. The separation between \\ these datasets is  determined to validate the user's identity. A variety of techniques, including \\ autoregressive analysis,  power spectral density, statistical histograms, and different combinations of \\ these features, are employed for user identification.\end{tabular} \\ \hline

\begin{tabular}[c]{@{}l@{}} \cite{hertweck2019brain} \\ \cite{weber2021structured}
\end{tabular}               & 2019                               & Brain Signals & BCI Headset                      & \begin{tabular}[c]{@{}l@{}}The dataset collected from the EEG-enabled HMD is influenced by the intervention of low-frequency \\ signals. Signal quality is assessed in this work, and a method for filtering out this noise is presented.\end{tabular}  \\ \hline

\begin{tabular}[c]{@{}l@{}} \cite{lu2018multifactor} \\ \cite{lu2019fmhash} \\ \cite{lu2020fmkit} \\ 
\cite{lu20213d} \\
\cite{lu2021global}
\end{tabular}   & 2018          & \begin{tabular}[c]{@{}c@{}}Air Handwriting \\  and Hand Geometry\end{tabular}    & Leap Motion Tracker             & \begin{tabular}[c]{@{}l@{}}
Determining whether a login request is challenging in terms of signal processing and matching, as it \\ is based on a piece of hand movement, which results in limited performance in existing systems. A \\ multifactor user authentication framework was proposed by the authors, which utilizes both the \\ motion signal of a piece of in-air handwriting and the geometry of the hand skeleton captured by a \\ depth camera.\end{tabular} \\ \hline

~\cite{li2017accurate, li2019brain}                           & 2017                               & Speech &  Blue Yeti Microphone                                & \begin{tabular}[c]{@{}l@{}}
Anti-eavesdropping password schemes are being focused on for development using smart glasses, \\ which are closely related to the VR headset device. Smart glasses come equipped with various \\ features, including gTapper, gRotator, and gTalker.\end{tabular}                \\ \hline

~\cite{shang2019enabling} & 2019 & Speech & Blue Yeti Microphone  &   \begin{tabular}[c]{@{}l@{}} Owing to the positioning of the microphone in the VR/AR headset, it is difficult to protect \\the VR system from the usual voice-based spoof attacks. Here, the internal body voice and \\ air-propagated  voice were used for spoof detection. The spectrograms from both channels are\\ generated, and then  their cross-correlation is computed. When both the channel voices belong to the \\ same user, the  correlation is high.  \end{tabular} \\ \hline

\end{tabular}}
\label{tab:BehaviorBio}
\end{table*}

\subsection{Behavioural biometric}

As discussed previously, knowledge-based biometrics are susceptible to attacks, highlighting the need for alternative approaches. One promising avenue is to leverage user behavior in response to stimuli or forced changes, as this behavior often contains identifiable information. Table \ref{tab:BehaviorBio} summarizes the behavioral biometrics that are extensively employed in XR systems for user verification.  

Finger-movement-based behavior biometrics was proposed by~\cite{liebers2021using} for reliable user authentication while using the XR system. Finger movement data were collected to identify users based on their interaction with buttons and sliders [\cite{liebers2022identifying}]. However, this method is limited in that it restricts the user behavior to specific input devices. Therefore, there is a need for an approach that can track the natural unrestricted movement of the hand. Hand movements are susceptible to mimicry attacks because they are exposed to the physical world [\cite{khan2018augmented, parampalli2008practical}].

Signatures have long been utilized as a traditional means of authentication, particularly in settings such as banks. However, signatures are vulnerable to forgery by trained individuals through practice. To address this limitation, a novel air handwriting-based verification system was proposed by~\cite{lu2018multifactor, lu2019fmhash, lu2020fmkit, lu20213d, lu2021global}. In this system, users use controllers to sign into the VR environment, ensuring that fraudulent actors do not gain access to their signatures. In addition, the system incorporates the tracking of hand orientation and geometry, providing a multi-factor authentication (MFA) mechanism for enhanced security. By capturing air signatures and monitoring hand orientation and geometry, this approach offers a secure and robust authentication method for VR environments. The combination of air signatures and multifactor authentication provides a comprehensive approach for user verification, minimizing the risk of unauthorized access.

Recent research by~\cite{virtualworlds1010004} has focused on identifying the minimal set of features necessary for user identification with the aim of creating a task-agnostic VR authentication system. Additionally, investigations have been conducted to determine the most effective machine-learning models for user identification using eye-gaze data. Notably, these studies predominantly relied on stored data for the analysis. In contrast, the study presented by~\cite{lohr2018implementation} proposed a real-time processing approach for eye movements in VR authentication. Researchers collect eye gaze data by tracking the movement of a sphere in 2D space, enabling the processing and analysis of eye gaze information in real time. The Gaze-based biometrics has also been developed for the authentication of XR devices [\cite{luo2020oculock}]. OcuLock [\cite{luo2020oculock}] system is based on gaze-based biometrics, which are secure and accurate. OcuLock utilizes multiple biometric modalities, including iris, eye movement, eyelid, extraocular cells, muscles, and surrounding nerves, to verify the identity of the user. Considering these various factors, OcuLock enhances the security and accuracy of biometric authentication in XR systems. In the context of VR authentication, eye movement- and gaze-based behavioral authentication methods are typically integrated into head-mounted displays. However, these approaches often require significant computational resources owing to the large number of eye-gaze features involved. Furthermore, these VR authentication systems are typically task dependent and identify users based on their performance in specific activities related to a particular task.

The use of full-body motion as a biometric for user verification has been explored by~\cite{prakash2018recent}. This approach can be considered as a multimodal approach for authentication, incorporating various behavioral biometrics. A related paper by~\cite{winkler2022questsim} presented a reinforcement learning framework that enables the real-time tracking of human body motion using sparse signals obtained from wearable devices. The proposed framework can simulate physically valid full-body motions based on input signals, even in scenarios where observations of the lower body are unavailable. Through the use of a policy network, the framework learns to generate appropriate torques for activities such as balancing, walking, and jogging while closely aligning with the input signals. The framework exhibits robustness across different locomotion styles, body sizes, and environments, demonstrating its effectiveness in immersive experiences such as augmented reality (AR) and virtual reality (VR). These movement-based behavioral biometrics, including full-body motion, have demonstrated promising results for user authentication. However, there remains a need for biometric approaches that are less susceptible to vulnerability and potential attacks. Further research and development is necessary to explore additional modalities and techniques that can enhance the security and reliability of user verification in immersive environments.

Behavioral biometrics based on brain and heart signals can be considered more reliable for biometric authentication in virtual reality (VR) and augmented reality (AR) systems (XR in general). Unlike other biometrics, brain signals such as electroencephalogram (EEG) signals are generated internally and collected through sensors on the head-mounted display (HMD) of the VR/AR device. In a particular study, EEG signals from two channels were collected: one while the user was in the virtual environment and the other while the user was looking at a laptop. The signals from both channels were compared using a Siamese network to compute distance. If the difference in distance was below a specific threshold, the user was considered authentic. However, the data-collection process using an HMD for brain-signal authentication can be tedious and inconvenient because of the large number of sensors involved. To address this inconvenience, another study by~\cite{hertweck2019brain, weber2021structured} determined the minimum number of sensors required for an accurate user authentication using brain signals. By identifying the bare minimum sensor configuration, this study aims to alleviate the handling and discomfort associated with HMD-based brain signal authentication. Similarly, electrocardiogram (ECG) signals, which capture heart activity, have been explored as less vulnerable biometrics for user authentication [\cite{sun2022perae}]. ECG-based authentication leverages the unique characteristics of an individual's heart signals for identification and verification.

Voice biometrics are widely used in XR environments to enhance the efficiency, accuracy, and robustness of user identification in XR environments. Voice-spoofing attacks pose a significant challenge in VR/AR systems, as fraudsters may attempt to gain unauthorized access to the device. While most voice spoofing work has focused on smartphone platforms, the special location of microphones in VR headsets adds complexity to spoof detection. In~\cite{shang2019enabling}, an internal body voice and an air-propagated voice were utilized for spoof detection. Spectrograms were generated from both channels and cross-correlation was computed. A high correlation indicated that both channels belonged to the same user, thus enhancing the security of voice authentication in the VR headsets. These approaches highlight the potential of utilizing internal body signals such as brain signals, heart signals, and voice characteristics to enhance the reliability and security of biometric authentication in VR/AR systems. Further research and development are needed to refine these techniques and overcome associated challenges.

\section{Digital avatars in XR}
\label{Avatars}
\begin{figure*}[htp]
\centering
\includegraphics[height= 300pt,width= 400pt ]{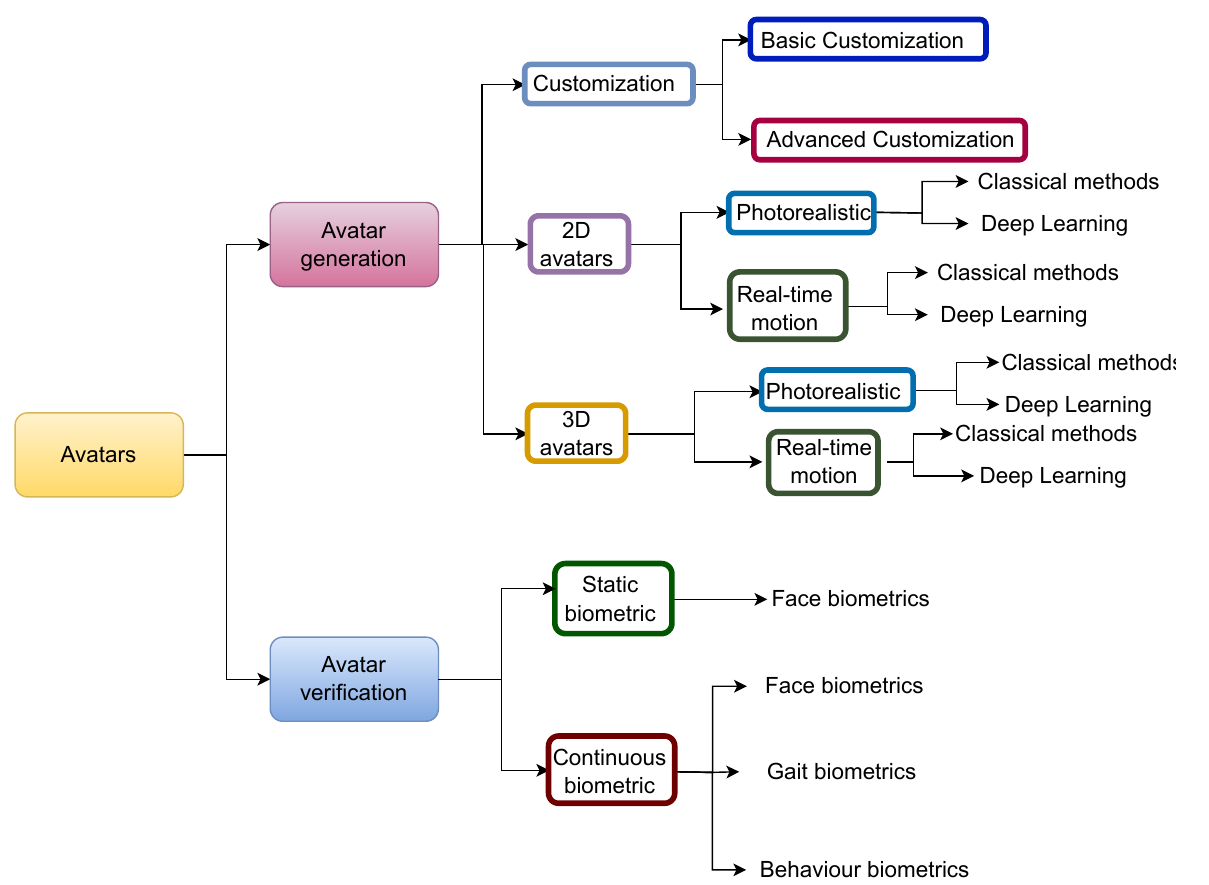}
\caption{Taxonomy representing of different techniques used for avatar generation and verification in XR scenario.}
\label{VR_taxonomy}
\end{figure*}

\begin{figure*}[htp]
\centering
\includegraphics[height= 200pt,width= 220pt ]{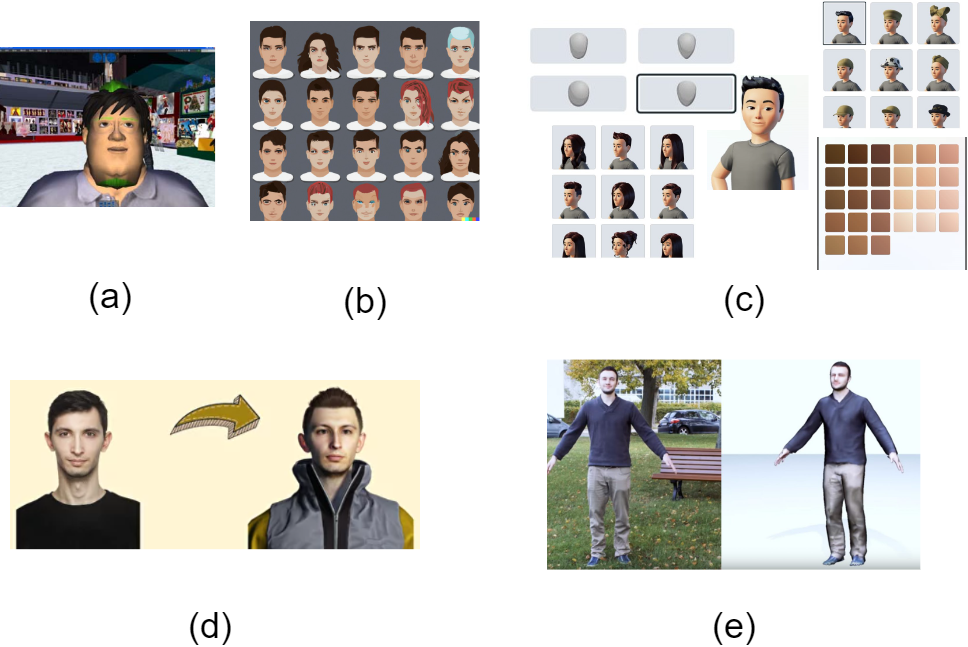}
\caption{Evolution of avatars over time for the use of virtual reality environments. (a) and (b) shows the basic customization, (c) has the advanced customization of hair, eye-color, face tone, etc. 2D and 3D photo-realistic avatars is shown in figure (d) and (e) displays the real-time motion of the photo-realistic avatars.}
\label{VR_avatars}
\end{figure*}

Digital avatars provide social interaction among users in an Extended Reality (XR) environment. Avatar has evolved over the years based on the advancements of avatar generation techniques and now can be used for verification of the individuals in the XR environment. The taxonomy of the current section is shown in figure~\ref{VR_taxonomy}. These avatars were generated in five different ways, as shown in Figure ~\ref{VR_avatars}.

\begin{itemize}
   \item  \textit{Early VR Systems:} In the early stages of VR, avatars are typically basic and lack detailed customization options. Users are often limited to selection from a small set of predefined avatars that resemble their real-world appearance [\cite{liu2021social}].   
   \item  \textit{Basic Customization}: As VR technology improves, users are given the ability to customize their avatars to some extent. They could choose from a range of pre-designed attributes such as gender, hairstyles, clothing, and basic facial features. Although this provided a degree of personalization, options were still relatively limited.    
   \item  \textit{Advanced Customization:} With the advancement of VR systems and increasing demand for immersive experiences, more comprehensive customization options have been introduced. Users can modify the finer details of their avatars, including facial structure, skin tone, body shape, and accessories. This level of customization allows users to create avatars that closely resemble their real-world appearance or desired virtual identity [\cite{o2016convergence,achenbach2017fast}].    
   \item \textit{Photorealistic Avatars:} The introduction of advanced capture technologies, such as 3D scanning and photogrammetry, revolutionized avatar generation. Users can now create highly realistic avatars by capturing their own facial features and bodies using specialized hardware or even standard cameras. These techniques enable the creation of avatars that closely mirror the user's physical appearance, thereby enhancing the sense of presence in VR environments.    
   \item  \textit{Real-time Motion Capture}: Recent developments in motion capture technologies have enabled real-time tracking and mapping of user movements on avatars. This allows for a more accurate representation of body movements, facial expressions, and gestures in a virtual environment. Users can see their avatars to replicate their real-world actions, thereby enhancing their immersion and sense of embodiment in VR [\cite{luginbuhl2011towards, feng2017just}]. Photo-realistic avatars serve as the identity of the user in the virtual world to uniquely identify the user.
\end{itemize}

% \subsection{Threats from digital avatars}

% Photo-realistic avatars, often generated by businesses and professionals to provide users with the best experience, contain a significant amount of personal information. However, these avatars can be susceptible to various attacks and misuse. Several attack vectors on digital avatars include identity infringement, anonymity misuse, and identity revelation based on avatar behavior.

% Identity infringement occurs when a fraudster impersonates a user, using a photo-realistic avatar, to deceive others and gain access to confidential information. Anonymization, which is commonly used to protect user identity, can also be misused if there is no effective tracking system to distinguish between genuine and fake anonymized individuals. 

% Virtual games that provide a platform where real people play are becoming very popular and can be accessed by all groups of people. These groups can range from small children and women to terrorist organizations. Studies show that terrorists are able to practice on simulators having an experience close to reality~\cite{o2007spies}. Therefore there is a need to identify the avatars in the virtual world

\subsection{Biometric-based photo-realistic avatars}

In this section, we summarise the existing works on biometrics based of Photo-realistic avatar generation on verification to achieve reliable and secure XR environment. First, we discuss the different avatar generations, including both 2D and 3D, then we discuss the improvents in the quality of avatar with the evolution of generative deep learning architectures, and finally, a  discussion on the avatar verification method is presented.   Table \ref{VR_avatars_Generation} shows the existing avatar generation using biometric characteristics.  

\subsubsection{Avatar generation:}

In the initial stages of Extended Reality (XR), avatars were relatively simplistic, offering users a limited choice. However, advancements in technology have led to the introduction of customizable avatars, allowing users to personalize their virtual representation by selecting features, such as hairstyle, eye color, and skin tone. Although these avatars offered a degree of customization, they were not realistic and were unsuitable for biometric purposes. Consequently, the demand for photorealistic avatars has increased in recent years. Researchers have conducted numerous studies on the generation of photorealistic avatars using both 2D and 3D techniques.

% \begin{table}[]
% \caption{2D photo-realistic avatar generation}
% \begin{tabular}{|l|l|l|}
% \hline
% \multicolumn{1}{|c|}{\textbf{Author}} & \multicolumn{1}{c|}{\textbf{Year}} & \multicolumn{1}{c|}{\textbf{Description}}                                 \\ \hline
% Kim et al.                            & 2000                               & \begin{tabular}[c]{@{}l@{}}The paper proposes a unified framework for real-time photo-realistic interactive\\ virtual environments using multiview cameras and stereo images/videos.\end{tabular} \\ \hline
% \end{tabular}
% \end{table}

\textbf{2D photo-realistic avatar:} The 2D photorealistic avatar has been extensively studied for face biometrics. The generated avatar possesses biometric characteristics. A unified framework for generating a photorealistic interactive virtual environment (piVE) without relying on bluescreen techniques or specialized rendering hardware has been proposed by~\cite{kim2000photorealistic}. The framework, as presented by~\cite{kim2000photorealistic}, utilizes heterogeneous multiview cameras and stereo images/videos to render piVE in real time. This approach eliminates the need for expensive high-end computers and renders graphics objects based on camera parameters and user interaction.

\begin{table*}[]
\centering
\caption{Existing techniques on Photo-realistic avatar generation that are widely used in XR scenario.}
\scalebox{0.7}{
\begin{tabular}{|l|l|l|l|}
\hline
\multicolumn{1}{|c|}{\textbf{Author}} & \multicolumn{1}{c|}{\textbf{Year}} & \textbf{Data Type} & \multicolumn{1}{c|}{\textbf{Description}}                                                       \\ \hline

~\cite{kim2000photorealistic}                            & 2000   & 2D                            & \begin{tabular}[c]{@{}l@{}}The paper proposes a unified framework for real-time photo-realistic interactive virtual environments \\ using multiview cameras and stereo images/videos.\end{tabular} \\ \hline

~\cite{lattas2023fitme}                       & 2023             & 3D                  & \begin{tabular}[c]{@{}l@{}}FitMe is a facial reflectance model and rendering pipeline that generates high-fidelity avatars \\ from single or  multiple images. It achieves photorealistic shading, preserves identity, and \\ outperforms existing methods. FitMe is efficient, producing relightable avatars in just one minute.\end{tabular} \\ \hline

 ~\cite{dong2023ag3d}                       & 2023             & 3D                  & \begin{tabular}[c]{@{}l@{}}
 Most of the current 3D avatar generation techniques rely on scarce 3D training data, making it expensive \\ and limited. In this work 3D images are created using abundant 2D images. by adversarial generative model.
 \end{tabular} \\ \hline

~\cite{szymanowicz2023photo}                       & 2023       & 3D                        & \begin{tabular}[c]{@{}l@{}}
PiCA is an efficient deep generative model for 3D human faces, using a convolutional architecture and \\ rendering adaptation. It achieves improved reconstruction and real-time rendering on mobile VR headsets.\end{tabular} \\ \hline

~\cite{szymanowicz2023photo}                       & 2023   & 3D                            & \begin{tabular}[c]{@{}l@{}}A landmark detector estimates camera poses from 360-degree videos, enabling immersive 3D experiences \\ and photo-realistic avatars using mobile phone cameras. Synthetic data validates its effectiveness.\end{tabular} \\ \hline

~\cite{szymanowicz2023photo}                       & 2022      & 3D                         & \begin{tabular}[c]{@{}l@{}}
The paper presents PointAvatar, a method for creating realistic and animatable head avatars from \\ casual video sequences. It uses a deformable point-based representation, separating color into albedo \\ and normal-dependent shading. PointAvatar outperforms existing methods in challenging scenarios \\ while being more training-efficient.\end{tabular} \\ \hline

~\cite{szymanowicz2023photo}                       & 2022      & 3D                         & \begin{tabular}[c]{@{}l@{}}
The study introduces a mobile-friendly method for generating realistic avatars with minimal data, \\ utilizing conditional representation and inverse rendering techniques. The resulting high-fidelity \\ avatars show improved visual quality and animation capabilities compared to existing lightweight \\ avatar creation methods..\end{tabular} \\ \hline

~\cite{prokudin2021smplpix}                       & 2021     & 3D                       & \begin{tabular}[c]{@{}l@{}}Deep generative models improve realism in synthetic human images, while a proposed network \\ converts 3D mesh vertices into photorealistic images, achieving superior photorealism and efficiency.\end{tabular} \\ \hline

~\cite{lombardi2018deep}                       & 2018      & 3D                      & \begin{tabular}[c]{@{}l@{}}A deep appearance model that combines facial geometry and appearance for realistic rendering. It uses \\ a data-driven approach with a deep variational autoencoder to model complex effects and correct imperfect \\ geometry, making it suitable for real-time interactive applications like Virtual Reality (VR).\end{tabular} \\ \hline

~\cite{tewari2017mofa}                       & 2017        & 3D                    & \begin{tabular}[c]{@{}l@{}}Introduces a model-based deep convolutional autoencoder that reconstructs 3D human faces from single \\ color images, achieving superior quality and representation through its combination of a CNN-based \\ encoder and a differentiable parametric decoder.\end{tabular} \\ \hline

~\cite{feng2017just}                       & 2017        & 3D                    & \begin{tabular}[c]{@{}l@{}}real-time avatars are generated in under 20 minutes through scanning and combining \\ body, face, and hand models. Photogrammetry, blend shape modeling, and rigging \\ techniques are employed for realistic facial expressions, hand shapes, and body movements\end{tabular} \\ \hline

~\cite{loper2015smpl}                       & 2015       & 3D                     & \begin{tabular}[c]{@{}l@{}}A network that combines geometry-based rendering with generative networks to generate photorealistic \\ images of human models without traditional rasterization. The model shows superior photorealism and \\ rendering efficiency compared to conventional renderers, addressing challenges in flexible control of the \\ generative process while preserving subject identity.

\end{tabular} \\ \hline

~\cite{michael2017model}                       & 2017       & Motion                     & \begin{tabular}[c]{@{}l@{}}
The paper proposes an automatic method for creating full-body avatars with minimal user intervention \\ and no requirement for a calibrated camera.

\end{tabular} \\ \hline

~\cite{sato2010using}                       & 2010       & Motion                     & \begin{tabular}[c]{@{}l@{}}The initial work for avatar generation is shown in which the motion capture is done by mechanical, \\optical, and magnetic sensing.

\end{tabular} \\ \hline

\end{tabular}}
\label{VR_avatars_Generation}
\end{table*}

\textbf{3D photo-realistic avatar:} The 3D photo-realistic avatar generation has gained the attention of researchers owing to its high-quality generation that can reflect the biometric characteristics of the individual. Recent advancements in deep generative models have significantly improved the realism of synthetic human images. However, flexibly controlling the generative process, such as changing camera angles and poses while preserving biometric identity features, remains a challenge. On the other hand, deformable body models such as SMPL [\cite{loper2015smpl}] provide pose and shape control, but rely on traditional rendering pipelines that have limitations. By~\cite{prokudin2021smplpix}, a network was proposed to bridge the gap between the geometry-based rendering and generative networks. The network directly converts a sparse set of 3D mesh vertices into photorealistic images, thereby eliminating the need for traditional rasterization. The model was trained on a large dataset of human 3D models and real photos, demonstrating superior photorealism and rendering efficiency compared to conventional differentiable renderers.

By~\cite{feng2017just}, the generation of real-time avatars was shown in less than $20$ min by scanning. The construction of a character involves the capture and creation of three distinct models: the body, face, and hands. A photogrammetric capture cage and software were used to create a 3D body model, which was then rigged to enable control over movement. To enable facial expressions and hand shapes, a blended shape model and hand rigging were integrated with the virtual avatar. A 3D face model was created by scanning facial expressions and constructing blended shapes. The face and body models were combined by replacing the face geometry and texture of the body model with the blend shape model, followed by color correction. Finally, a controllable hand and finger model were added to the 3D body model. In [\cite{tewari2017mofa}], a model-based deep convolutional autoencoder for reconstructing 3D human faces from single in-the-wild color images. The model combines a convolutional encoder network with a differentiable parametric decoder that encodes the detailed face pose, shape, expression, skin reflectance, and scene illumination. This allows the CNN-based encoder to extract meaningful parameters from a monocular image, thereby enabling end-to-end unsupervised training on large real-world datasets. Reconstructed faces demonstrated superior quality and representation.

The deep appearance model for a realistic rendering of the human face was presented by~\cite{lombardi2018deep} which utilizes a data-driven pipeline that learns a joint representation of facial geometry and appearance from multiview captures.  A deep variational autoencoder is employed to model vertex positions and view-specific textures, enabling the capture of complex effects and correction for imperfect geometry. The system can be implemented in real-time interactive settings such as Virtual Reality (VR) using existing rendering engines. This offers a flexible and accurate approach without the need for precise geometric estimates.

In [\cite{cao2022authentic}] photorealistic avatars of individuals were generated using a simplified mobile phone capture. Unlike other existing methods, the architecture employed in [\cite{cao2022authentic}] focuses on producing avatars that closely resemble individuals, while requiring minimal data. The approach involves utilizing a conditional representation and a universal avatar prior trained on the facial performance of a wide range of subjects. By employing inverse rendering techniques, the model was fine-tuned to enhance realism and personalize the avatar's range of motion. The technique proposed by~\cite{cao2022authentic} can create a high-fidelity 3D head avatar that accurately depicts the facial shape and appearance of subjects, along with providing control mechanisms for gaze direction. Experimental results demonstrate that the avatars generated using this approach exhibit superior visual quality compared to existing methods [\cite{wang2023swiftavatar}] for lightweight avatar creation.

The generation of realistic head avatars from video sequences has been proposed in [\cite{Zheng2022pointavatar}].  Existing methods rely on either fixed 3D morphable meshes (3DMM) or neural implicit representations, both of which have limitations in terms of topology, deformation, and rendering efficiency. In contrast, the approach in [\cite{Zheng2022pointavatar}] referred to as PointAvatar, introduces a deformable point-based representation that separates the source color into intrinsic albedo and normal-dependent shading. PointAvatar [\cite{Zheng2022pointavatar}] combines high-quality geometry and appearance with topological flexibility, ease of deformation, and rendering efficiency. Therefore, it can successfully generate animatable 3D avatars using monocular videos from various sources, surpassing other existing approaches in challenging scenarios, such as thin hair strands, while also being more training-efficient than competing methods.

To enable immersive 3D experiences and create photo-realistic avatars using commodity hardware, such as mobile phone cameras, a novel landmark detector was proposed by~\cite{szymanowicz2023photo}. It estimates the camera poses from 360-degree videos of a human head, allowing realistic rendering from any viewpoint. Synthetic data validation experiments demonstrate the effectiveness of the method, which was showcased with 360-degree avatars trained from mobile phone videos. In~\cite{lattas2023fitme}, FitMe, a facial reflectance model, and a rendering pipeline for generating high-fidelity human avatars from single or multiple images were presented. It utilizes a multimodal style-based generator and PCA-based shape model to capture facial appearance and shape. The differentiable rendering optimization process achieves photorealistic shading and accurately preserves identity. FitMe outperforms existing methods in reflectance acquisition and identity preservation by generating relightable avatars from single "in-the-wild" images or multiple unconstrained images of the same person. Notably, FitMe is efficient, requiring only one minute to produce avatars with a mesh and texture suitable for end-user applications. 

A computationally efficient and adaptive deep generative model for 3D human faces, called Pixel Codec Avatars (PiCA), was introduced by ~\cite{ma2021pixel}. PiCA combines a fully convolutional architecture for spatially varying feature decoding with a rendering-adaptive perpixel decoder. The model was trained in a weakly supervised manner and achieved improved reconstruction performance across different expressions, views, genders, and skin tones. PiCA is significantly smaller than existing models and enables the real-time rendering of multiple avatars in a single scene on a mobile VR headset.

The paper by~\cite{dong2023ag3d} discusses the challenges in creating 3D avatars from 2D images. Most of the current methods often rely on scarce 3D training data, making it expensive and limited. The proposed solution in~\cite{dong2023ag3d} aims to generate realistic 3D avatars from abundant 2D images by employing an adversarial generative model. This model captures body shape, clothing deformation, and articulation through a holistic 3D generator and flexible articulation module. To enhance realism, they use multiple discriminators and integrate 2D normal maps. Experimental results demonstrate that their approach surpasses previous methods in terms of geometry and appearance, validated through systematic ablation studies, highlighting the method's effectiveness and component significance.

\textbf{Real-time motion capture:} The goal of real-time motion capture is to render a full-body avatar with photorealistic quality to capture full-body biometric features. The initial work that captured the motion to generate avatars was presented in~\cite{sato2010using}. A full-body avatar requires capturing the motions using mechanical, optical, and electromagnetic waves. Mechanical motion capture utilizes rigid metal parts worn by an actor, with sensors that detect movements. It offers advantages such as no interference from magnetic or light fields, and certain movement detection. However, it lacks knowledge of body orientation unless supplemented with other sensors. Optical motion capture involves retroreflective markers tracked by infrared cameras. These benefits include freedom of movement, larger recording spaces, and detailed data. Drawbacks include occlusion, discomfort from wearing markers, and high costs. The initial work reported in~\cite{sato2010using} utilized an optical system with 12 cameras for a comprehensive body capture. Electromagnetic motion capture relies on magnetic receivers worn by the actor and lacks blind spots and relative affordability. Disadvantages include distortion caused by metals and restricted movement due to cable connections.

\begin{figure*}[htp]
\centering
\includegraphics[height= 225pt,width= 360pt ]{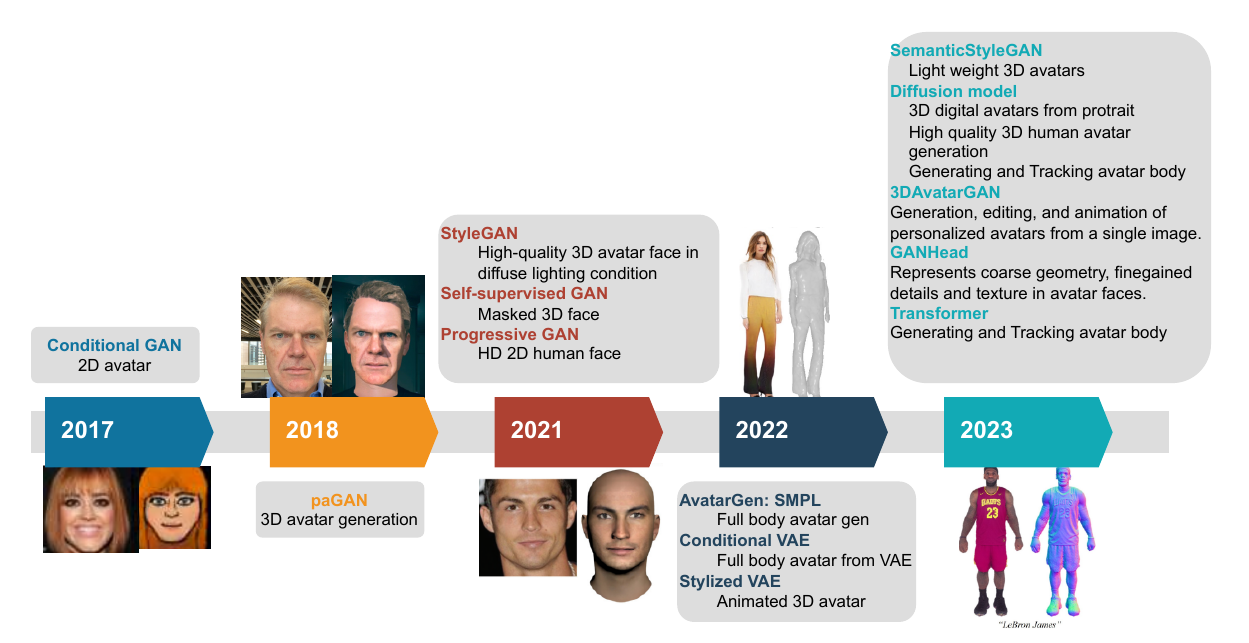}
\caption{The timeline of the development of avatars through generative deep learning models.}
\label{AvatarDLModel}
\end{figure*}

In~\cite{michael2017model}, an automatic method for creating full-body avatars with minimal user intervention and no requirement for a calibrated camera was presented. The technique involves training a 3D shape model using PCA on full-body 3D scans, followed by training a multiview Active Shape Model (ASM) using 2D silhouettes. Uncalibrated cameras capture orthogonal input images, and the silhouettes are refined and registered using the ASM. Facial features were detected, and a 3D avatar was generated by optimizing the shape parameters and mapping textures. Regressor models control the body parameters, enabling the creation of a random template population. Templates undergo skeletonization, rigging, and animation, resulting in animation-ready avatars for VR integration.

Recently,  Facebook Reality Labs (FRL) have made significant progress in creating photo-realistic avatars with Codec Avatar technology. To further enhance realism and create full-body avatars, the FRL has established two high-quality capture studios: one for capturing facial details and another for full-body motion. These studios are equipped with numerous high-resolution cameras that capture data at an impressive rate of 1 GB per second. This extensive camera setup pushes the limits of their capture hardware but enables the collection of superior data to create highly realistic avatars. Additionally, one of the studios incorporates 1,700 microphones to reconstruct immersive 3D sound fields, crucial for an immersive virtual environment\footnote{https://tech.facebook.com/reality-labs/2019/3/codec-avatars-facebook-reality-labs/}.

\textbf{Summary of photorealistic avatar generation over the years:} The timeline of avatar generation has seen remarkable progress, evolving alongside advancements in deep learning models. As depicted in Figure~\ref{AvatarDLModel}, the journey began with the introduction and utilization of generative deep learning architectures such as Generative Adversarial Networks (GANs), Variational Autoencoders (VAEs), and diffusion techniques. One notable milestone was reached in 2017 with   successfully generated 2D avatars closely resembling human faces through Conditional GANs [\cite{wolf2017unsupervised}]. However, these avatars fell short in meeting the requirements for biometric tasks.

In 2018 with the introduction of paGAN, as outlined in~\cite{nagano2018pagan}. This approach marked a significant improvement by emphasizing facial biometric characteristics and adopting an end-to-end deep learning strategy for synthesizing facial textures. Subsequent years, particularly in 2021, witnessed a leap towards photorealistic avatars. Architectures such as StyleGAN [\cite{luo2021normalized}], self-supervised GAN [\cite{lu2021general}], and progressive GAN [\cite{li2021animated}] have played pivotal roles in achieving this shift. Photo-realistic avatars not only excel in capturing facial biometric features but also showcase finer textures, contributing to enhanced realism.

While photo-realistic avatars have excelled in facial biometrics, the incorporation of gait movements for stronger and more secure authentication was done in 2022. Two key models, the Conditional VAE [\cite{aliakbarian2022flag}] and stylized VAE [\cite{sang2022agileavatar}], were introduced to generate realistic gait movements. The focus on gait movements continued to gain momentum, and in 2023, better results were achieved with the use of the diffusion [\cite{huang2023humannorm}] and transformer [\cite{aliakbarian2023hmd}] models. These models seamlessly integrate facial information and gait movements, resulting in avatars that offer a more robust and secure authentication experience. The constant evolution in deep learning techniques has not only shaped the field of avatar generation, but also holds promising implications for the future of biometric authentication.

%%%%%%%%%%%%%%%%%%%%%%%%%

% The drawback of this approach is that we need multiview cameras to generate photo-realistic avatars.

\begin{table*}[]
\centering
\caption{Existing techniques on Avatar verification techniques in XR scenario.}
\scalebox{0.75}{
\begin{tabular}{|l|l|l|l|}
\hline
\multicolumn{1}{|c|}{\textbf{Author}} & \multicolumn{1}{c|}{\textbf{Year}} & \multicolumn{1}{c|}{\textbf{Biometric Modality}} & \multicolumn{1}{c|}{\textbf{Description}}                                                                                    
\\ \hline                                            

~\cite{gavrilova2011applying} & 2011 & Face & \begin{tabular}[c]{@{}l@{}}
This paper discusses the advancements in virtual reality security, \\ emphasizing avatar authentication methodologies and their connection \\ to biometric research
\end{tabular}

\\ \hline

~\cite{ajina2011evaluation}                          & 2011                               & Face          & \begin{tabular}[c]{@{}l@{}}
Avatar faces are characterized using the wavelet features and a support \\ vector machine (SVM) classifier is used for the classification task.
\end{tabular} \\ \hline

~\cite{yampolskiy2012face}                          & 2012                               & Face + Behaviour                                  & \begin{tabular}[c]{@{}l@{}}Proposed an algorithm to track the person in both the real and virtual\\  world. The matching is done for the faces of human-vs-avatar, avatar-vs\\-avatar, avatars from inter VR world, and avatar vs sketch made by humans.\end{tabular} \\ \hline

~\cite{yampolskiy2012experiments}                          & 2012                               & Behaviour and attributes                       & \begin{tabular}[c]{@{}l@{}}Proposed an algorithm to track the person in the virtual world. The\\ identification is done using the attributes, behavioral patterns, and apperance.\end{tabular} \\ \hline

~\cite{thies2016facevr} & 2016 & Face &
\begin{tabular}[c]{@{}l@{}}This paper presents a face-motion capture of photorealistic avatars \\ and compares it with the motion of a real face motion to verify the user.\end{tabular}
\\ \hline

~\cite{sethuraman2023metasecure} & 2023 & Face &
\begin{tabular}[c]{@{}l@{}}This paper presents a method for safeguarding identity in the metaverse by \\ combining PIN/password authentication with avatar recognition.\end{tabular}
\\ \hline

\end{tabular}}
\label{VR_avatars_Verfi}
\end{table*}

\subsubsection{Avatar verification}

This section discusses existing methods that are employed to verify photo-realistic avatars of users. Avatar verification plays an important role in ensuring that a legitimate user has gained access to the system in both physical and virtual worlds. Table \ref{VR_avatars_Verfi} shows the existing works on Avatar verification. 

%This section emphasizes the importance of authentication for photo-realistic avatars, particularly during the generation stage, to address the security threats they pose. Section~\ref{BiometricVR} discusses various modes of biometric authentication in VR, primarily focused on user access to virtual reality devices. However, there is a possibility that fraudsters may bypass device authentication, gain unauthorized access to the VR device, and create digital avatars.

To mitigate the risks associated with digital avatars, authentication must be implemented during the creation of photorealistic avatars and continuously tracked in the XR space. This stage is crucial because once the avatar is generated, it becomes challenging to identify the user based solely on the avatar's appearance. 
Early work on avatar verification was presented in~\cite{gavrilova2011applying} where the static features from the avatar were verified based on landmark points.  The use of more sophisticated features such as wavelets and an SVM classifier was proposed by~\cite{ajina2011evaluation} to achieve reliable avatar verification in enrolment conditions. The use of motion-based cues, such as eye gaze movement, was proposed in~\cite{thies2016facevr} which can verify a photorealistic avatar in the XR space. The use of motion-based cues can enable the generalizable verification of avatars with different rendering qualities. To further enhance security, a new password-free system through a three-tier approach was proposed by~\cite{sethuraman2023metasecure}. It relies on verifying the legitimacy of the device being used, recognizing the user's face, and employing physical security tools such as security keys, all in compliance with Fast Identity Online (FIDO2) standards.

Continuous evaluation methods such as user identification through speech, brain signals, or other biometric modalities can be utilized to ensure both verification and liveness detection. The real-time behavior of avatars in virtual environment [\cite{yampolskiy2012experiments}] based on movement patterns, interaction preferences, body language, facial expressions, and gestures specific to an individual person is used in the behavior attributes to verify the avatar. These behavioral traits are captured and analyzed in real-time to authenticate the user or to identify them within the virtual environment. In ~\cite{yampolskiy2012face}, the fusion of face and behavior cues was presented to further enhance avatar verification, which can be extended to both static and continuous avatar verification. The advantage of using behavioral biometrics in virtual reality is that they are difficult to fake or replicate because they are inherent to a user's natural behavior. This makes them a potentially effective means of authentication or identification within a virtual environment.

By implementing authentication measures during avatar generation and incorporating continuous evaluation methods, the security and integrity of digital avatars can be enhanced, preventing unauthorized access and misuse of these representations.

\section{Biometric databases in XR}
\label{Databases}
In this section, we describe physiological and behavioral biometric datasets reflecting XR environments, which are widely used by researchers to develop biometric verification algorithms in XR environments. Furthermore, we discuss the face avatar datasets used for the biometric verification. 

\subsection{Physiological biometric datasets}

Table \ref{Tab:DB_Phy} summarizes the physiological biometric datasets widely employed for verification in XR applications.  The LPW dataset [\cite{tonsen2016labelled}] is a pupil detection dataset collected in the form of a video. There were $66$ high-quality recording samples collected from $22$ participants using an HMD. The dataset is diverse in the form of indoor-outdoor illumination and participants wearing spectacles, makeup, and contact lenses. For image processing, video samples must be converted into images that require extra processing. In~\cite{kim2019nvgaze}, the NVGaze dataset was released, which contained approximately 2 million synthetic and 2.5 million real images of 35 participants. The images cover a wide range of modularities, such as pupil, iris, sclera, skin tone, face shape, and gaze direction. OpenEDS is another large-scale image dataset that covers modalities such as the iris, sclera, and pupil. The advantage of the OpenEDS dataset is that it contains pixel-level annotations of 12,759 images. In addition, there were 252,690 unlabelled and 91200 video frames. In~\cite{kotwal2024vrbiom}, VRBiom dataset, a periocular dataset for biometrics in XR domain is released. 900 short videos of 10 seconds length from 25 individuals is captured  using  internal tracking cameras of Meta Quest Pro at 72 FPS. This dataset is the first of its kind to publicly feature realistic, non-frontal views of the periocular region, designed for various biometric applications.

The The Brain signals were used for biometrics captured from the electrodes on the HMD. In~\cite{li2019brain}, EEG signal data were collected from 32 volunteers of both sexes. Data were collected in VR and non-VR scenarios for 180 s. For the first 30 s, the participant was asked to keep their eyes closed, and in the second 30 s, eyes were opened. The video was played in the final 120 s. From the collected EEG signals, temporal and spectral features can be extracted, and various machine-learning classifiers can be used to verify the speaker. To further level up user involvement in data collection, task-based VR was used to collect data [\cite{gregory2022dataset}]. EEG data were collected from 47 participants. The participants were presented with a scenario in which a glass was present at a location in a VR environment that may be half or fully filled. The user must remember the location and status of the glasses. The participant had to tell the location and status of the glass with the help of a cue (avatar or stick). EEG data were collected at multiple instants during the task. EEG was recorded when (1) the cue appeared, (2) the glass appeared, (3) the user responded to the location, and (4) the user responded to the status. To verify that user features were extracted from the collected EEG data, classifiers were used.

\begin{table*}[]
\centering
\caption{Physiological biometrics datasets employed by researchers addressing biometric verification in  XR applications}
\scalebox{0.85}{
\begin{tabular}{|l|l|l|l|l|l|}
\hline
\textbf{Author} & \textbf{Year} & \textbf{Biometric Modality} & \textbf{Type} & \textbf{\#Samples} & \textbf{\#Participants}\\ \hline
~\cite{tonsen2016labelled}       & 2016          & Pupil detection             & Video & \begin{tabular}[c]{@{}l@{}}   66 video+\\130,856 video frames \end{tabular} 
& 22                                                   \\ \hline
~\cite{kim2019nvgaze}       & 2019          & \begin{tabular}[c]{@{}l@{}}Iris, pupil, skin-tone\\ and face shape\end{tabular} & Image & \begin{tabular}[c]{@{}l@{}} 2 million synthetic+\\2.5 million real  \end{tabular} & 35 \\ \hline
~\cite{garbin2020dataset,garbin2019openeds} & 2020          & Iris+sclera+pupil    & Image+Video & \begin{tabular}[c]{@{}l@{}}12,759 labelled+\\ 252,690 unlabeled+\\ 91200 frames from video\end{tabular} &  152 \\ \hline

~\cite{kotwal2024vrbiom} & 2024          & Periocular    & Video & 900 short videos &  25 \\ \hline

~\cite{li2019brain}  & 2019          & Brain signals             & EEG & 256 & 32                                                  \\ \hline

~\cite{gregory2022dataset}          & 2022          & Brain signals         & EEG signal & 10387 & 47                                                         \\ \hline
~\cite{sun2022perae}            & 2022          & Heart signal           & ECG   & - & -                                                      \\ \hline
~\cite{krishna2019multimodal}   & 2019          &   
\begin{tabular}[c]{@{}l@{}}   Brain signals+\\Eye tracking\end{tabular} 
& EEG+Image & - & -                                                \\ \hline
~\cite{tran2023anti}     & 2022          & Finger vein             &  Image& \begin{tabular}[c]{@{}l@{}} 1476~\cite{asaari2014fusion} \\ 3816~\cite{yin2011sdumla} \\ 1220~\cite{yang2014database}   \end{tabular} 

&   \begin{tabular}[c]{@{}l@{}}   123~\cite{asaari2014fusion}\\106~\cite{yin2011sdumla}\\610~\cite{yang2014database}\end{tabular}

\\ \hline

\end{tabular}}
\label{Tab:DB_Phy}
\end{table*}

\subsection{Behavioral biometric datasets}

Table \ref{Tab:DB_Bhev} summarizes the behavioral biometrics datasets widely employed in XR applications.  The behavioral datasets were constructed  based on tracking the user's head, hand, eye, and entire body movements. In~\cite{friedman2017method}, a gait and eye movement dataset was proposed that contained images collected from 1000 participants.  In~\cite{griffith2021gazebase}, a monocular dataset was collected from 322 college participants, consist of 12334 samples.  Every participant was recorded six times while performing seven eye-tracing tasks. (1) Forcing task in which the pupil movement is recorded, (2) tracking the horizontal movement of the black spot, (3) tracking of eye movement when the participant views the 3D video, (4) recording the eye movement when the user performs the reading task on the text displayed on the screen in the VR environment, (5) tracking the eye movement when the black sphere jumps from one position to another in the VR environment; (6) Fixation task in which the participant is said to fix eye movement on the target without blinking, and (7) the participant is said to play a game in VR in which there are red-blue balls and they have to remove a ball of a particular color by performing fixation on the ball they want to remove. In~\cite{lohr2023gazebasevr}, the previous dataset is extended from a monocular to the binocular eye-tracking dataset is collected that comprises 5020 samples from 407 college participants. Every participant was recorded six times while performing five eye-tracing tasks. They were said to have performed the first five out of 7 from the monocular dataset. The eye-movement information is presented in the ($x,y,z$) coordinate format with a timestamp of nanosecond precision.

\begin{table*}[htp]
\centering
%\resizebox{1\columnwidth}{!}{%
\caption{Behavioral  biometrics datasets  employed by researchers addressing biometric verification in  XR applications}
\scalebox{0.95}{
\begin{tabular}{|l|l|l|l|l|l|}
\hline
\textbf{Author} & \textbf{Year} & \textbf{Biometric Modality} & \textbf{Type} & \textbf{\#Samples} & \textbf{\#Participants}
\\ \hline
~\cite{sun2022perae}            & 2023          & Hand Movement           &   Image & 4,520,828 & 15   
\\ \hline
~\cite{lohr2023gazebasevr}            & 2023          & GazeBaseVR    & Video & 5,020 & 407   \\ \hline
~\cite{aziz2022synchroneyes}                & 2022          & Eye movement & Video & &     \\ \hline
~\cite{griffith2021gazebase}            & 2021          & Gaze Base & Video & 12,334 & 322       \\ \hline

~\cite{lohr2020eye} & 2020 & Eye movement & Image & - & 468
\\ \hline
~\cite{friedman2017method} & 2017 & Gait+EyeMovemet & Image & - & 1000
\\ \hline
~\cite{sun2022perae}            & 2023          & Hand Movement           &   Image & 4,520,828 & 15                                                      \\ \hline
\end{tabular}}
%}
\label{Tab:DB_Bhev}
\end{table*}

%%%%%%%
\subsection{Facial Avatar datasets used in XR applications}
Table \ref{Tab:DB_Avtar} summarizes the facial avatar datasets widely employed in XR applications for biometric verification.  Various datasets that have been used in the biometrics of avatars and to demonstrate presentation attacks and detection are available in the literature. With the use of deep learning techniques, such as variation autoencoders (VAE) and generative adversarial networks (GAN), it has become possible to generate photo-realistic avatars. In~\cite{oursler2009parameterized}, ten images per avatar were created. The avatar images were obtained from various camera angles. In~\cite{tewari2017mofa}, the 3D human face is generated from a single image of the human face. In~\cite{lombardi2018deep}, the VAE was used for the generation of avatars, in which various finer intricacies such as oral cavities and eyelashes were also addressed. This results in  avatars that are heavy in size and number of pixels. Pixal codec avatars are lightweight and decode visible pixels for real-time creation of avatars [\cite{ma2021pixel}]. 

\begin{table}[]
\centering
\caption{Avatar recognition dataset in XR applications}
\begin{tabular}{|l|l|l|l|l|l|}
\hline
\textbf{Author} & \textbf{Year} & \textbf{Type} & \textbf{\#Samples} & \textbf{\#Avatars}\\ \hline
~\cite{ma2021pixel}            & 2021              & Face & - & -  \\ \hline

~\cite{bader2014sid}                & 2014           & Entire body& 350 & 50    \\ \hline

~\cite{ajina2011evaluation}            & 2011           & Face & 1800 & 100       \\ \hline

~\cite{oursler2009parameterized} & 2009 & Face & 10/avatar & -

\\ \hline

\end{tabular}
\label{Tab:DB_Avtar}
\end{table}

\section{Performance evaluation metrics}
\label{PerformanceMetrics}
In this section, we discuss the performance metrics used to benchmark biometric performance in XR applications.  Performance metrics are critical for evaluating the effectiveness and efficiency of biometric authentication systems for XR. These metrics provide a quantitative analysis of the accuracy and reliability of biometric authentication systems and help identify areas that require improvement. The following metrics are widely employed to benchmark biometric verification in XR.

\begin{itemize}
    \item \textbf{False Acceptance Rate (FAR):} FAR measures the percentage of times the system incorrectly identifies an unauthorized user as an authorized user. This metric is important for determining the accuracy of biometric authentication methods, such as facial recognition and voice recognition. In virtual reality systems, FAR can help determine the system's accuracy in identifying users and preventing unauthorized access.

    \item \textbf{False Rejection Rate (FRR):} FRR measures the percentage of times the system incorrectly rejects an authorized user. This metric is crucial for determining the effectiveness of biometric authentication methods. In virtual reality systems, FRR can help determine the system's ability to accurately recognize and authenticate authorized users.

    \item \textbf{Equal error rate (EER):}  EER [\cite{cheng2004method}] is the point at which the False Acceptance Rate (FAR) and False Rejection Rate (FRR) are equal. This metric is used to measure the overall accuracy of the biometric authentication system. In virtual reality systems, EER can help determine the overall accuracy and effectiveness of the system in identifying and authenticating users.

    \item \textbf{Accuracy:} Accuracy measures the number of correct or authentic users when compared with the total number of users. In virtual reality systems, accuracy can help determine the overall effectiveness of a biometric authentication system in identifying and authenticating users.
    
\end{itemize}

\section{Discussion in Research Questions}
\label{Sec:QA}

\begin{figure*}[htp]
\centering
\includegraphics[height= 270pt,width= 320pt]{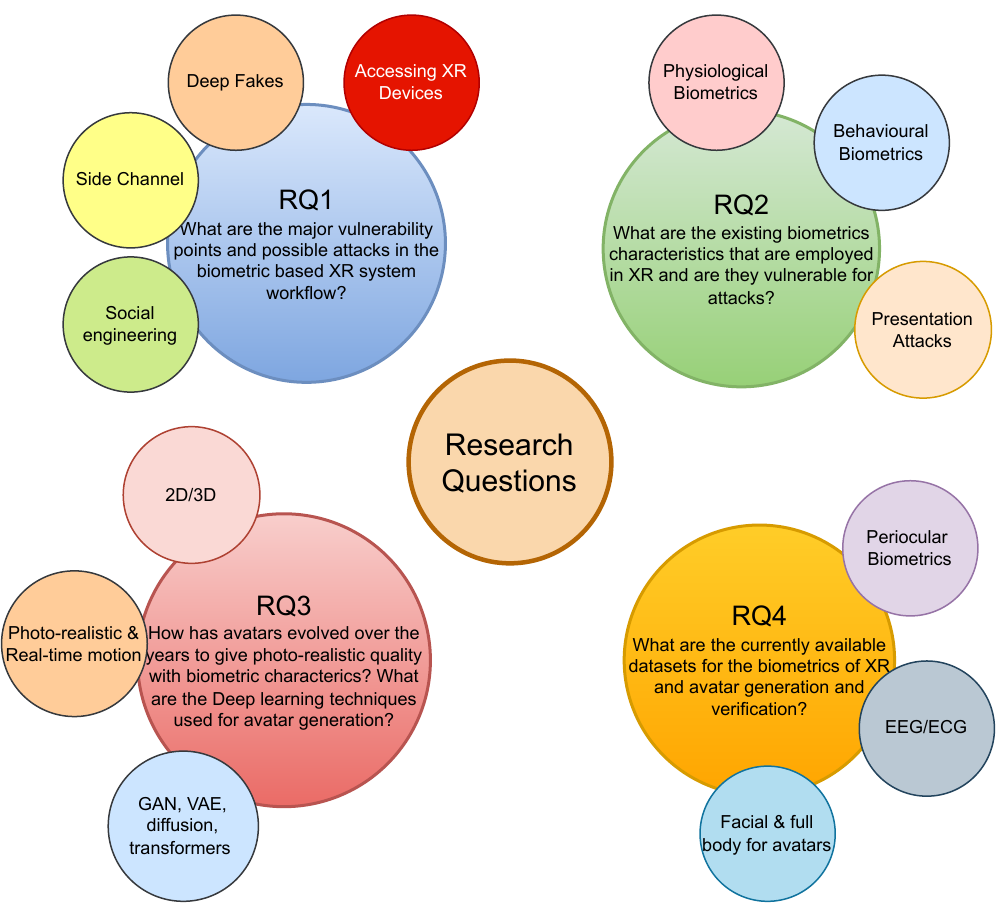}
\caption{Key answers to the proposed research questions}
\label{RQ_RR}
\end{figure*}

In Section \ref{Sec:Intro}, we introduce four key inquiries pertaining to the utilization of biometrics in XR systems. These inquiries served as the foundation for literature review. Through a comprehensive analysis of pertinent articles, we have now attained the ability to respond to the aforementioned questions, as delineated below and in figure~\ref{RQ_RR}.

\begin{itemize}
    \item \textbf{RQ1:} \textbf{What are the major vulnerability points and possible attacks in the biometric based XR system workflow?} 
During our literature review and examination of XR system workflows, we identified three primary points of vulnerability related to the use of biometrics in XR. The first vulnerability arises during the initial authentication, when a user gains access to the device for the first time. This vulnerability is exploited through presentation attacks and injecting deep fakes, which represent targets involving biometric characteristics. Second, attacks may occur through side-channel methods, where hand gestures and body movements can be intercepted by hidden cameras and microphones within the HMD. Finally, vulnerabilities may manifest when the user is immersed in an XR environment. In such scenarios, attackers infiltrate the XR world through social engineering techniques, accessing and exploiting body movement data utilized for authentication within the XR environment.

    \item \textbf{RQ2:} \textbf{What are the existing biometrics characteristics that are employed in XR and are they vulnerable for attacks?}

Physiological and behavioral biometrics were extensively investigated, as detailed in Section ~\ref{BiometricVR}. Physiological biometrics related to fingerveins and iris have demonstrated promising security in the XR domain. Behavioral biometrics encompasses the collection of user data pertaining to external body motions, including those of the eyes, fingers, hands, head, and full body, as well as internal responses such as speech and brain signals. Fingerveins are situated beneath the skin, rendering them less susceptible to presentation attacks, whereas the iris is more vulnerable to such attacks.  Similarly, for behavioral biometrics, a combination of information regarding hand and head orientation and movement is utilized to enhance security.

Existing VR and AR devices use biometrics as their primary authentication component. Apple's vision pro~\footnote{https://www.apple.com/apple-vision-pro/} uses OpticID, which captures detailed iris data of the user. The privacy of OpticID is ensured by encrypting and protecting it using a secure enclave. Iris data never leaves the user's device. Other devices like Vive Pro Eye~\footnote{https://www.vive.com/sea/product/vive-pro-eye/overview/}, Varjo~\footnote{https://varjo.com/products/xr-4/}, and Neo 2 Eye~\footnote{https://www.tobii.com/products/integration/xr-headsets/device-integrations/pico-neo-2-eye} also utilize iris  as biometrics.
    
    \item \textbf{RQ3:}  \textbf{How has avatars evolved over the years to give photo-realistic quality with biometric characterics? What are the Deep learning techniques used for avatar generation?}

Avatars have undergone significant evolution over the years. In the early stages of virtual reality (VR), avatars are characterized by predefined and customized features that often do not accurately represent biometric characteristics. However, with advancements in deep learning techniques, avatars have transitioned to photorealistic representations and real-time motion simulations. This transformation has been facilitated by Generative Adversarial Network (GAN)-based deep-learning architectures, enabling the generation of photorealistic 3D avatars. Furthermore, the utilization of a Variational Autoencoder (VAE), diffusion, and transformer-based models has enabled the creation of full-body avatars capable of capturing both physiological and behavioral characteristics with remarkable fidelity.
    
    \item \textbf{RQ4:} \textbf{What are the currently available datasets for the biometrics of XR and avatar generation and verification?}
Section~\ref{Databases} describes existing datasets for both physiological and behavioral biometrics. For external biometrics, the modes of data collection are primarily image, video, and speech. For behavior data collection, such as the brain and heart, information is collected using EEG and ECG. The data were mainly collected from the age group of 20-30 years old data subjects.

Facial data were collected primarily to generate photorealistic avatars. There are also a few works that have now started generating a full-body avatar dedicated to biometrics. There are very few datasets dedicated to the biometric study of avatars, and with the advancements in deep learning architectures, there should be more collection of photorealistic and full-body motion avatars.    
\end{itemize}

\section{Open challenges}
\label{OpenChallenges}
In this section, we discuss the open challenges in the area of biometric applications in XR that are essential to address to achieve reliable and trustworthy biometric verification while using XR. The following are the important challenges: 

\begin{itemize}
 \item \textbf{Devices that can collect multiple modalities of biometrics from the same XR device:}
Biometric authentication has rapidly gained popularity as a reliable method for verifying the identities of individuals. It involves using unique physical or behavioral characteristics, such as iris, speech, brain signal, and gaze, to identify and authenticate an individual. These modalities contain complementary and unique information about the user, which can be leveraged to provide stronger authentication systems. The use of biometric authentication in XR is not new, but current XR devices are limited to capturing only one modality. This is where multi-model XR biometrics are developed, which combines multiple biometric modalities to create a robust authentication system. Multimodal biometrics in the XR technology have significant advantages over single-modal biometrics. By combining multiple biometric modalities, the authentication system becomes more secure and challenging for attackers. This is because it is more difficult to fake multiple biometric traits than a single trait. Additionally, multimodal biometric systems are more accurate, as they rely on more than one form of biometric data, reducing the chances of false positives or negatives. The development of multi-model XR biometrics is still in its early stages, and further research is required to optimize this technology. The complexity of integrating multiple biometric modalities presents a challenge, particularly with respect to real-time data processing.

\item \textbf{Need to accommodate the users of various age groups in XR applications:}
Extended Reality (XR) technology is designed to be used by all age groups, from young children to the elderly. With the increasing use of XR technology, there is a growing need for robust biometrics that can provide security. However, most XR-based datasets used for biometric authentication are collected from college students, which raises concerns about the validity and applicability of such datasets to other age groups. The biometric traits of different age groups are known to be different, which can have a significant impact on the accuracy and effectiveness of biometric authentication in XR. For instance, the iris recognition system may not be as reliable for young children whose irises are still developing or for the elderly whose iris texture may change due to age-related conditions such as cataracts. Therefore, it is essential to analyze diverse data in terms of age groups to ensure that biometric authentication systems are reliable for all age groups. This involves collecting and analyzing biometric data from a diverse range of age groups, and using these data to train and test the biometric authentication system.

Furthermore, it is crucial to address the issue of bias in biometric datasets, which may result in inaccurate authentication for certain age groups or ethnicity's. Bias can occur if the dataset used to train the biometric authentication system is not diverse enough or if there is a lack of representation of certain age groups or ethnicity's. Therefore, it is important to ensure that the datasets used for biometric authentication are diverse and inclusive and that the analysis is performed on data that are representative of the entire population.

\item \textbf{The study of possible presentation attacks specific to the XR biometric systems:}
As extended reality has become more integrated into our daily lives, the security of biometric authentication in extended reality systems has become increasingly crucial. Although extensive research has been conducted on presentation attacks in various biometric modalities, such as speech [\cite{wu2015spoofing}], iris [\cite{czajka2018presentation}], and face recognition [\cite{ramachandra2017presentation}], little attention has been paid to how these attacks can be executed on biometric authentication systems within the XR framework. For instance, in the case of speech biometrics, attackers can attempt to replicate a user's voice using text-to-speech (TTS) or voice conversion techniques to replay the recorded audio. Similarly, iris recognition systems can be fooled by presenting a fake iris image or a contact lens with the same iris pattern. Face-recognition systems can be tricked by presenting a 3D-printed mask or deep-fake video. The rapid advancement in deepfake technologies, especially in generating hyper-realistic images and videos, poses significant security challenges for Extended Reality (XR) systems. As these systems heavily rely on visual and sensory data to create immersive environments, they become increasingly vulnerable to manipulation through deepfakes. Malicious actors could inject false visuals, alter user interactions, or manipulate virtual environments, leading to misinformation, identity theft, and other cyber threats. Open research in this field focuses on developing robust detection algorithms, improving authentication techniques, and designing countermeasures that can safeguard XR systems from such deepfake attacks while maintaining user trust and experience.

However, it is unclear how these attacks operate at ground level within the context of an extended reality environment. Would these attacks be more or less effective in an extended reality system compared to a traditional system? What are the unique challenges and vulnerabilities of biometric authentication systems in extended reality and how can they be addressed? As extended reality becomes more prevalent in areas such as healthcare, education, and finance, it is crucial to address these questions and develop robust biometric authentication systems that are resistant to presentation attacks in an extended reality environment. This requires collaboration between researchers, VR developers, and security experts to identify potential threats and develop countermeasures to protect the security and privacy of VR users.

\item \textbf{Selection of biometric characteristics  based on the XR application:}
The extended Reality (XR) technology is now extensively used in various fields and requires a suitable and efficient biometric authentication system. The selection of biometric characteristics should be based on the required level of security, cost of implementation, and user experience. High-cost biometric systems, such as iris, retina scans, voiceprints, and brain signals, are suitable for high-security tasks such as financial transactions or national security discussions. However, the same biometric system may not be feasible for activities such as gaming or virtual tourism, where the user experience is the priority. Low-cost biometric systems, such as hand gesture recognition or eye tracking, and face biometrics may be suitable for such activities.

%It is important to note that a single biometric system may not provide sufficient security. A combination of biometric systems can be used to increase security and reduce the chance of successful attacks. For example, a combination of face and voice recognition can be used for banking transactions in a XR environment. This can make it difficult for attackers to breach the system because multiple biometric factors are required for authentication. However, the privacy of the users must be addressed before integrating the biometrics. 

\item \textbf{Performance metrics tailored to XR:}
XR biometric authentication systems are a subset of biometric authentication systems used in a simulated environment that require different performance metrics compared to traditional biometric systems. Although EER and accuracy are commonly used metrics in traditional biometric systems, they may not be appropriate for XR biometric authentication systems. This is because XR biometric authentication systems are subject to unique challenges such as virtual environment distortion, hardware limitations, and user experience factors. For example, XR headsets can cause discomfort or may not fit properly, resulting in poor data collection and incorrect authentication. Additionally, XR biometric authentication systems require real-time response and accuracy, which can be difficult to achieve because of hardware limitations. Therefore, it is important to develop performance metrics specific to XR biometric authentication systems. For example, one such metric could be the "immersion score," which measures the level of immersion the user experiences in the virtual environment during the authentication process. Another metric could be the "latency score," which would measure the time taken to authenticate the user in a real-time environment. The development of specific performance metrics for XR biometric authentication systems, including immersion score, latency, and EER, would help evaluate the effectiveness and efficiency of these systems more accurately. This would enable developers to identify weaknesses in the system and improve user experience. Ultimately, this will lead to a more reliable and secure XR biometric authentication system.

\item \textbf{Continuous authentication of user with XR devices:}
Biometric authentication is a popular method for verifying the identity of a user before granting access to a device or a system. However, once a user is authenticated and logged in, there is a risk that someone else can take over a session if the user leaves the system unattended or if the user's credentials are compromised. This is the point where the concept of continuous authentication occurs. In the context of extended reality, continuous authentication is particularly important owing to the immersive nature of the experience. Users can easily lose track of time and forget that they are in an extended reality environment, making them more vulnerable to attacks. For example, a user may unknowingly wander into an unauthorized area of a virtual world or they may leave their session unattended, giving an attacker the opportunity to take over. Continuous authentication can also be used to detect suspicious behaviors in real time. For example, if the system detects that a user is behaving erratically or is attempting to access unauthorized areas, it can alert appropriate authorities or take other measures to prevent a potential security breach.

\item \textbf{Continuous verification of avatar in virtual space:}
Avatars are digital representations of the users in virtual reality systems. They offer a unique opportunity to customize the appearance of the user and create a personalized experience. However, the potential for avatar misuse by cyber criminals cannot be ignored. Fraudsters can easily impersonate others by creating an avatar with someone else's identity and using it for malicious activities such as committing cyber crimes, harassment, or identity theft. This can cause serious harm to both the victim and reputation of the virtual reality system.

To address this challenge, it is necessary to have a robust avatar authentication system that can verify a user's identity and continuously monitor their behavior in a virtual environment. The authentication system can use various biometric modalities, such as facial, voice, and fingervein recognition, to verify the user's identity before creating an avatar. Additionally, it is important to track the user's behavior continuously to ensure that the avatar is used only by the authorized user. One approach to continuously monitor a user's behavior is to detect anomalous behavior and flag it for further investigation. For example, if an avatar is used in an unauthorized location or if the user's behavior suddenly changes, the system can trigger an alert to the system administrator or security personnel. The system can also use a combination of biometric and behavioral modalities to increase the accuracy of avatar authentication and reduce the risk of fraudulent activities.

    % \item Testbeds
    
\item \textbf{Need for standardization forum:}
There is a pressing need for a standardized forum that establishes guidelines and benchmarks for extended reality (XR) biometrics. Currently, various datasets are inadequate for conducting comprehensive commercial-level tests. However, these datasets may not accurately reflect real-world scenarios or encompass a diverse range of users and environments. Consequently, it is crucial to develop a standardized dataset that can serve as a benchmark for evaluating the biometric features of XR, ensuring their readiness for commercial purposes. A key challenge in XR biometrics is the vulnerability of certain biometric features to presentation attacks and other forms of exploitation. Presentation attacks involve the use of fake or altered biometric traits to deceive a biometric system, potentially granting unauthorized access.

The establishment of a standardized forum would involve collaboration between industry experts, researchers, and policymakers. This forum would work towards defining the criteria for a robust and reliable XR biometric system. The criteria would include aspects such as the quality of biometric data acquisition, diversity of the dataset, and resilience of the biometric features against presentation attacks. Establishing a standardized benchmark would enable developers and manufacturers to effectively evaluate and enhance their XR biometric systems.  

\item \textbf{Usability of XR:} The usability of extended reality (XR) technologies, including virtual reality (VR) and augmented reality (AR), faces several challenges. One key challenge is the privacy and security concerns associated with the use of biometric data in XR applications. Although biometric authentication methods such as facial recognition can enhance security, usability, and convenience, there is a risk of unauthorized access and data breaches, raising significant privacy concerns. Ensuring the reliability of biometric systems is another challenge because inaccurate recognition algorithms can lead to authentication failures. Apple's Optic ID~\footnote{https://www.apple.com/apple-vision-pro/} system for the Apple Vision Pro headset takes steps to address these challenges by focusing on eye recognition, which offers a high level of security and accuracy. It protects privacy by restricting app access to eye data, and employs 3D information to prevent spoofing. User data is encrypted and stored locally, accessible only to a secure enclave processor, enhancing privacy and security~\footnote{https://www.pandasecurity.com/en/mediacenter/technology/apple-vision-pros-privacy-security/}. However, further challenges remain, such as ensuring the continued security of biometric data storage and addressing the potential vulnerabilities that could emerge as technology evolves. Additionally, ongoing efforts are needed to educate users about the benefits and risks of biometrics in XR, and to establish industry standards for responsible biometric data handling.
  
\end{itemize}

\section{Conclusion}
\label{Conclusion}
Biometric applications with extended reality systems are gaining popularity owing to their diverse applications in education, e-commerce, entertainment, and healthcare. This paper presents a comprehensive overview of biometric authentication methods that employ both physiological and behavioral biometrics. An exclusive discussion of the different vulnerability points in a biometric-based XR  system is presented for the first time. Furthermore, this study explores security aspects by discussing physiological and behavioral biometric techniques for XR. It covers the generation and verification of photorealistic 2D and 3D avatars and touches upon databases containing physiological, behavioral biometric data, and avatar datasets. The study also introduces performance metrics for assessing XR authentication systems and concludes by addressing the current challenges and potential solutions to XR authentication methods.

\section{Appendix}

\begin{itemize}

    \item XR - Extended Reality
    \item VR - Virtual Realilty
    \item AR - Augmented Reality
    \item HMD - Head Mounted Display    
    \item MFA - Multi Factor Authentication
    \item PAI -  Presentation Attack Instruments
    \item EEG - Electroencephalogram
    \item ECG - Electro cardiogram
    \item DFT - Discrete Fourier Transform
    \item FFT - Fast Fourier Transform
    \item OM - Ordinal Measure
    \item PSD - Power Spectral Density
    \item WPT - Wavelet Packet Transform    
    \item BCI -  Brain-Computer Interface
    \item SVM - Support Vector Machine
    \item PCA - Principal Component Analysis
    \item GRU -  Gated Recurrent Unit
    \item GAN - Generative Adversarial Network
    \item VAE - Variational Auto-encoder
    \item CNN - Convolutional Neural Network
    \item PiCA - Pixel Codec Avatars.
    \item EER - Equal Error Rate
    
\end{itemize}

\bibliographystyle{unsrtnat}

% \bibliography{sn-bibliography}  

%%% Uncomment this line and comment out the ``thebibliography'' section below to use the external .bib file (using bibtex) .





% \end{thebibliography}

\end{document}